\pdfoutput=1
\documentclass[letterpaper, 10 pt, journal, twoside]{IEEEtran} 

\usepackage{times}
\IEEEoverridecommandlockouts

\usepackage{verbatim}
\usepackage{amsmath}
\usepackage{algorithm}
\usepackage[noend]{algpseudocode}
\usepackage{graphicx}
\usepackage{amssymb}
\usepackage{amsmath}
\usepackage{amsthm}
\usepackage{array}
\usepackage[draft]{fixme}
\usepackage[english]{babel}
\usepackage{verbatim}
\usepackage{parskip}
\usepackage{tikz}
\usepackage{multirow}
\usepackage{xspace}
\usepackage{color}
\usepackage{tikz}
\usepackage{caption}
\usepackage{subcaption}
\usepackage[colorinlistoftodos]{todonotes}
\usepackage[font=small,labelfont=bf,tableposition=top]{caption}
\usepackage{empheq}
\usepackage{cases}
\usepackage[normalem]{ulem}
\newcommand\LyonsStrikeout{\bgroup\markoverwith
{\textcolor{cyan}{\rule[0.5ex]{2pt}{0.8pt}}}\ULon}
\usepackage{bm}
\newcommand{\lly}[1]{\textcolor{cyan}{ \bf #1}\color{black}}
\makeatletter
\def\BState{\State\hskip-\ALG@thistlm}
\makeatother                           

\DeclareCaptionLabelFormat{andtable}{#1~#2  \&  \tablename~\thetable}

\usepackage{booktabs, tabularx}
\usepackage{multirow}

\usepackage{xspace}

\theoremstyle{definition}
\newtheorem{example}{Example}

\newtheorem{problem}{Problem}

\theoremstyle{definition}
\newtheorem{assumption}{Assumption}

\newcommand{\manu}[1]{\textcolor{blue}{ \bf #1}\color{black}}

\DeclareMathOperator*{\argmax}{arg\,max}

\title{Distributed Attack-Resilient Platooning Against False Data Injection}

\author{{Lorenzo Lyons, Manuel Boldrer, Laura Ferranti} 
  \thanks{L. Lyons, L. Ferranti are with the Dept. of Cognitive Robotics, Delft University of Technology, Mekelweg 2,
    2628 CD, Delft, Netherlands, {\tt
      \{l.lyons, l.ferranti\}@tudelft.nl}, M. Boldrer is with the Dept. of Cybernetics, Czech Technical University, Karlovo namesti 13, 12135, Prague, Czechia, {\tt 
       boldrman@fel.cvut.cz}.}
      \thanks{This research is supported by the NWO-TTW Veni project
HARMONIA (no. 18165).}}
\begin{document}

\maketitle

\begin{abstract}
This paper presents a novel distributed vehicle platooning control and coordination strategy. We propose a distributed predecessor-follower CACC scheme that allows to choose an arbitrarily small inter-vehicle distance while guaranteeing no rear-end collisions occur, even in the presence of undetected cyber-attacks on the communication channels such as false data injection. The safety guarantees of the CACC policy are derived by combining a sensor-based ACC policy that explicitly accounts for actuator saturation, and a communication-based predictive term that has state-dependent limits on its control authority, thus containing the effects of an unreliable communication channel. An undetected attack may still however be able to degrade platooning performance. To mitigate it, we propose a tailored Kalman observer-based attack detection algorithm that initially triggers a switch from the CACC policy to the ACC policy. Subsequently, by relying on a high-level coordinator, our strategy allows to isolate a compromised vehicle from the platoon formation by reconfiguring the platoon topology itself. The coordinator can also handle merging and splitting requests. We compare our algorithm in an extensive simulation study against a state of the art distributed MPC scheme and a robust control scheme. We additionally extensively test our full method in practice on a real system, a team of scaled-down car-like robots. Furthermore, we share the code to run both the simulations and robotic experiments.

\end{abstract}

{\small{\textbf{\textit{Video---}{https://youtu.be/M0ukDXu6Mk0}}}}

{\small{\textbf{\textit{Github---}{https://github.com/Lorenzo-Lyons/Distributed-Attack-Resilient-Platooning-Against-False-Data-Injection}}}}


\IEEEpeerreviewmaketitle

\section{Introduction}
\label{sec:Introduction}

Autonomous vehicles are becoming a reality and promise to radically change the world of transportation. In the last decade, many studies have been focused on platooning of autonomous vehicles. Platooning driving involves a group of vehicles traveling in a formation, with each vehicle maintaining a constant distance behind the preceding vehicle. Platoon-based driving can decrease traffic congestion, reduce emissions, improve road safety and driving comfort~\cite{hall2005vehicle}, as well as alleviate the need for human drivers, that is becoming an even more pressing issue~\cite{driver_shortage}.

To enhance coordination among vehicles in a platoon, it is essential to combine the use of on-board sensors and a communication network. Relying solely on on-board sensors can limit the vehicles' understanding of their neighbors' intended motion, affecting the overall platoon performance (e.g., more conservative inter-vehicle margins). While a communication network allows vehicles to exchange information about their intentions and improve platoon performance, communication among the vehicles can also introduce vulnerabilities into the system. In fact, the information transmitted through a communication network is more susceptible to malicious attacks, such as false data injection or denial of service~\cite{ahmad2018man}.

\begin{figure}
        \centering
        \includegraphics[width=\columnwidth]{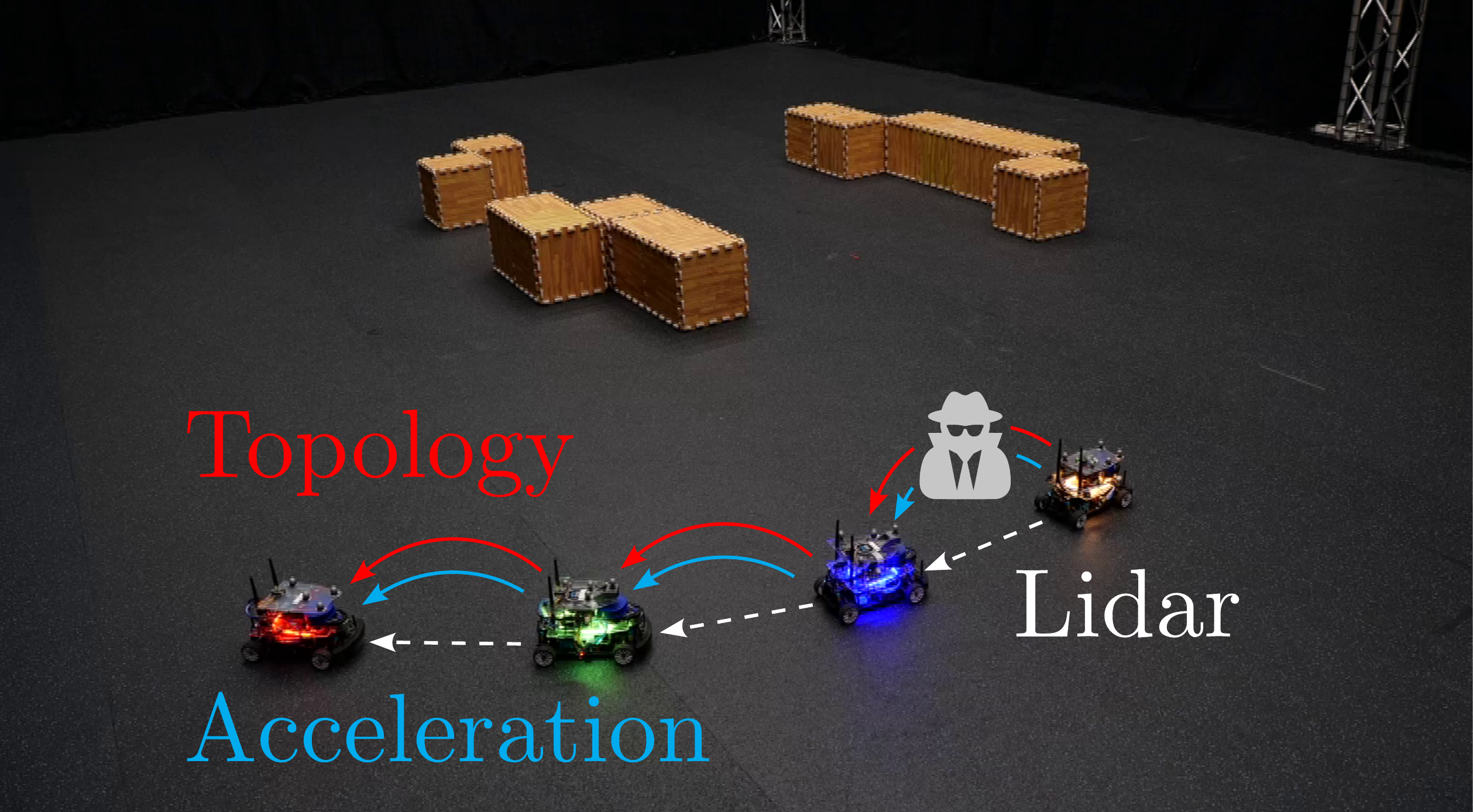}
        \caption{Experimental platform consisting of 4 scaled-down car-like robots~\cite{DART}. Each robot measures its state relative to its predecessor using a Lidar and receives acceleration and platoon topology data through the communication network. Data flowing through the network is however susceptible to man-in-the-middle attacks (e.g. between the leading vehicle and its follower as shown in the figure) that we mitigate with our strategy.}
        \label{fig:jetracer intro picture}
    
\end{figure}

This paper focuses on the design of a distributed algorithm to achieve attack-resilient longitudinal platooning. The algorithm combines on-board sensing and vehicle-to-vehicle (V2V) communication. It provides collision avoidance guarantees and is also able to detect and isolate malicious attacks to continue the operation of the platoon by conveniently reconfiguring the platoon formation. It can furthermore manage merging and splitting requests. This work considers false data injection attacks on the acceleration signal that is sent from the predecessor to follower vehicle, (see Figure~\ref{fig:jetracer intro picture}). We consider the case where the attacker may have full knowledge of the system and acts in a strategic manner to cause 
a crash in the platoon.\looseness=-1

\subsection{Related works}
Vehicle platooning is largely addressed in the literature. As~\cite{lesch2021overview} describes, platooning can be addressed on two different levels, \emph{Platooning control}, which relates to the design of the low-level controller that allows to maintain the desired distance among the vehicles and is responsible for the safety and stability of the formation, and \emph{Platooning coordination}, which takes care of the platoon management, that is, the organization and composition of the platoon formation itself.
\subsubsection{Platooning control} There is a plethora of methods in the literature. Traditional approaches rely on linear feedback controllers such as~\cite{de1999stability}, while in recent years model predictive controller (MPC) based solutions, such as~\cite{zheng2016distributed,thormann2020safe,basiri2020distributed,gratzer2022string}, have become increasingly popular. 
Strategies based on techniques other than feedback controllers or MPC include artificial potential fields~\cite{semsar2016cooperative}, consensus-based platooning~\cite{santini2016consensus}, and Deep Reinforcement Learning ~\cite{ferdowsi2018robust}.

Irrespectively of the specific platooning strategy, a necessary requirement for real-world applicability is providing safety by avoiding rear-end collisions. In the literature, this requirement has almost always been translated mathematically into the problem of proving string stability. Although definitions may vary slightly, the core idea is that disturbances in acceleration, velocity, or position introduced at the head of the platoon should not be amplified as they propagate downstream. In this work, we formalize string stability using a frequency-domain transfer function approach (see Section~\ref{sec:platoon_controller}), where we require that the magnitude of the transfer function from one vehicle to the next remains below one across all frequencies. This ensures that oscillations decay along the platoon. However, as we show, string stability alone is not sufficient to guarantee collision avoidance in the presence of actuator saturation, which we address through additional safety constraints.

In Adaptive Cruise Control (ACC) based on traditional linear feedback controllers, the typical way to prove string stability is to rely on frequency domain analysis and show that the magnitude of input and state signals is decreasing along the platoon~\cite{de1999stability}. However, since these methods can not include actuator limitations, collisions in the head of the platoon may still occur during emergency braking scenarios as shown in ~\cite{de1999stability}, where the issue of safety is addressed by performing an \textit{a posteriori} analysis that identifies the set of safe controller parameters, yet no \textit{a priori} safety guarantees or design criteria are provided.



When a platooning strategy uses communication among vehicles it is referred to as Cooperative-ACC (CACC). This brings significant advantages, since vehicles can now act in a predictive way by knowing their predecessor's intentions. The most popular class of CACC strategies is MPC since it also allows to consider actuation limits. The connection between actuator limits and safety concerns is further detailed in~\cite{tudelft_bidirectional_string_stab_and_act_bounds}, where collision avoidance is provided in a bidirectional cooperative setting by ensuring that saturation will not be reached in nominal platooning conditions, while in~\cite{MPC_with_governor_actuation_limits} the authors rely on an a worst-case preceding vehicle prediction to evaluate the risk of collision. 

However, the benefits of communication-based platooning strategies can be completely compromised in the presence of malicious attacks. As argued in~~\cite{ahmad2018man}, the risk of an adversarial attack on wireless communication is far greater than an attack on the on-board sensors, which makes cyber-physical security a major concern for CACC. The consequences of a malicious attack can be catastrophic, since, once injected, the false information may flow in the network, affecting the whole system and threatening efficiency, stability, and safety of the formation. The authors of~\cite{van2017analyzing} reviewed the impact of several attacks on existing platoon controllers.
In the literature, there are different approaches to deal with malicious attacks, as is discussed in~\cite{zhou2022robust}, but all of them suffer from specific issues.

Observer-based control strategies~(e.g., \cite{jahanshahi2018attack}) may experience latency problems due to the detection and mitigation mechanism. Robust control strategies (e.g., \cite{zheng2017platooning}) can generate high control inputs, which can be unfeasible in a real scenario due to input saturation. Adaptive  control strategies (e.g.,~\cite{jin2019adaptive}) can experience unwanted transient response issues. MPC strategies do not easily provide guarantees on the existence of feasible control solutions. 

Another specific issue with MPC that often eludes the literature's attention is that after a vehicle detects a compromised communication channel it can not simply shut it down, since the MPC requires the preceding vehicle's predicted trajectory as a reference. To circumvent the issue attack-resilient strategies either rely on identifying the attack vector and reconstructing the true signal~\cite{yang2024decoupling}, or generating a worst case reference trajectory for the preceding vehicle~\cite{ferrari_recent}. These solutions however either suffer from latency issues because they rely on a bank of past data to build a prediction of the preceding vehicle's behavior or may result in overly conservative inter-vehicle margins.

A promising approach is presented in~\cite{switching_CACC_ACC_delays} and~\cite{switching_cacc_acc_false_data_MPC}, where the authors leverage the assumption that on-board sensors are more reliable than the communication channels and define an MPC-based CACC policy that will switch to a sensor-based ACC strategy in the presence of either significant communication time delays as in~\cite{switching_CACC_ACC_delays}, or when a malicious attack is detected as in~\cite{switching_cacc_acc_false_data_MPC}. 

However we identify 3 issues with these approaches. Firstly, although these strategies can disable a compromised communication link and revert to an ACC policy, the downstream vehicles are now left without an open-loop reference for the MPC. Secondly, in the case of a malicious attack,~\cite{switching_cacc_acc_false_data_MPC} relies on a quick attack detection module to avoid safety issues during the time when a malicious attack is present but not yet detected. Lastly, when switching from CACC to ACC mode there is a discontinuity in the control strategy and guaranteeing safety after the ACC takes over is not trivial in the presence of actuator saturation.

\subsubsection{Platooning coordination} Platooning coordination is less addressed in the literature~\cite{lesch2021overview}. A fuel-optimal centralized solution is proposed in~\cite{van2015fuel}. In~\cite{johansson2018multi} the authors propose a non-cooperative game that models the multi-fleet platoon matching problem. In~\cite{de2019bio} a decentralized bio-inspired strategy for platoon management is proposed. A taxonomy on the objectives that may be considered for platooning is provided in~\cite{sturm2020taxonomy}. In~\cite{krupitzer2018towards} the authors present infrastructure-aided platoon management strategies for highways and urban areas, while in~\cite{sacone2021centralized} the authors propose both a centralized and decentralized high level platoon coordinator, which updates the parameters of the low-level platoon controller based on traffic conditions. While most platoon coordination methods focus on traffic flow management and managing platoon merging and splitting requests, our proposed coordinator is specifically designed to mitigate the effects of a malicious attack on the communication channels. In particular, leveraging the predecessor-follower topology to isolate a compromised vehicle. Notice that in practice, an attacker may be a vehicle affected by malicious software or by a man-in-the-middle attack at the communication network level~\cite{ahmad2018man}.
We furthermore observed a lack of literature related to malicious attacks at the coordination level, this is mainly due to the fact that the proposed strategies are mostly centralized or decentralized. In contrast, we propose a distributed solution that can also deal with the presence of an attacker in the network. We differentiate between attacks at the platoon coordinator level from attacks on the platoon controller according to the type of data being compromised. Attacks at the coordinator level concern topology information, i.e. in what sequence the vehicles should be arranged. Attacks on the platoon controller concern data about the vehicle's state such as acceleration and braking. Typically the latter is more safety critical since it's the controller's function to maintain the specified distance between vehicles, while the former deals with higher level decision making about the platoon formation.

\subsection{Our contribution and paper organization} 

This paper presents a novel distributed and attack-resilient algorithm for platooning.
It features three main contributions. First, 
we provide a low-level platoon controller that combines a linear sensor-based ACC controller, which is unaffected by network corruption, with a communication-based CACC controller, which improves performance but may be vulnerable to false data injection (FDI) in the communication channel. The proposed strategy falls in the category of ACC-CACC switching controllers. Differently from~\cite{switching_cacc_acc_false_data_MPC}, we do not rely on MPC, since when a vehicle switches to ACC mode, it will not have an open loop prediction of its intended motion to pass to its following vehicle. The proposed method addresses the limitations we identified in the state of the art:
\begin{itemize}
    \item In~\cite{switching_cacc_acc_false_data_MPC} the authors rely on the implicit assumption that a malicious attack will be detected quickly enough to prevent any safety-related concerns, while our method is able to guarantee safety even if a severe attack remains undetected indefinitely. To do so, similarly to~\cite{MPC_with_governor_actuation_limits}, we define a set of constraints that limit the authority of the communication-based controller, i.e. a \emph{safety filter}, and perform a worst-case scenario reachability analysis. In contrast to~\cite{MPC_with_governor_actuation_limits}, we additionally consider communication and design the safety filter to keep the vehicle state within a safe set even in case of a malicious attack.
    
    \item Concerning the fallback ACC algorithm, we directly address the issue of guaranteeing safety when switching to ACC mode of operation. We design a gain tuning procedure
    that provides both string stability and collision avoidance at the head of the platoon by considering actuation limits. Furthermore, the gain tuning procedure allows one to select an arbitrarily small inter-vehicle distance. This is of particular relevance since the highest gain in terms of aerodynamic drag and fuel efficiency are achieved for very small inter-vehicle distances of around $5-10$m \cite{fuel_savings}, and typical constant time-headway policies (such as~\cite{de1999stability}) do not allow to select a fixed spacing among vehicles, or lack formal \textit{a priori} safety guarantees.

\end{itemize}

Second, we propose a novel distributed platoon coordination strategy that allows to reorganize the platoon in order to isolate compromised vehicles, further limiting the effect of a malicious attack. The coordinator module, that runs on each vehicle, is also able to handle platoon merging and splitting requests. 
In the literature, only centralized or decentralized strategies are proposed for this purpose (not distributed). Moreover, our approach directly considers the reliability of a communication channel as a parameter to reorganize the platoon. Furthermore, our platoon coordinator can detect and isolate a malicious vehicle that communicates false data at the coordinator level. 

Third, we implemented the proposed platooning algorithm on real robotic platforms using a team of 4 scaled-down car-like robots~\cite{DART}, see Figure~\ref{fig:jetracer intro picture}, and provide a working code base to deploy our method in practice~\cite{GitRepo}. The videos of the experiments can be found in~\cite{video}.

To complete our framework we also present a tailored Kalman filter-based attack-detection policy specifically designed for platooning applications, that signals both to the low-level platoon controller to switch from CACC to ACC mode, and to the platoon coordinator to trigger a platoon reconfiguration maneuver.

The paper is organized as follows. Sec. II introduces the problem. Sec. III provides the proposed algorithm for resilient platooning. Sec. IV shows the simulation comparison between our low-level platoon controller, the MPC strategy~\cite{zheng2016distributed} and the robust control approach~\cite{kafash2018constraining}. Sec. V shows the experimental results. Sec. VI concludes the paper.


\section{Problem description}
\label{sec:Background}


This paper proposes a distributed solution for attack-resilient platooning. This problem is formulated as
follows:

\begin{problem}
\label{pr:problem_formulation}
Given $N$ vehicles, we want to achieve platoon-based driving, where each vehicle maintains a constant distance from the preceding vehicle. The platoon-driving condition has to:
\begin{enumerate}
    \item ensure \textit{safety}, that is, avoid inter-vehicle collisions;
    \item ensure (predecessor-follower) \emph{longitudinal string stability}. A system is \textit{string stable} if any error in position, velocity or acceleration does not amplify along the platoon~\cite{de1999stability};
    \item mitigate \textit{malicious attacks on the V2V communication}, that is, if a vehicle in the platoon communicates false acceleration data to its follower, the system must preserve safety and string stability; 
    \item enable \textit{dynamic platoon configuration}, that is, the system must manage changes in the platoon configuration, such as merging, splitting and reorganization.     \item \textit{isolate} the effects of a compromised vehicle once an attack has been detected.   
\end{enumerate}
\end{problem}
In addition, in the remainder of the paper, we make the following assumptions:
\begin{assumption}[\emph{On-board sensors}]
\label{as:assumption1}
 The sensors provide to the $i$-th vehicle its own velocity, and the position and velocity of the preceding vehicle. As motivated in section~\ref{sec:attack feasibility}, we consider the sensors to provide reliable state measurements.
\end{assumption}
\begin{assumption}[\textit{Graph topologies}]
\label{as:assumption2}
For the design of the platoon controller, we rely on the predecessor follower topology, while for the platoon coordinator, we assume a bidirectional predecessor follower topology (see Figure~\ref{fig:network}). We adopt a predecessor-follower topology in the platoon controller in order to provide stronger safety guarantees, as detailed in section~\ref{sec:platoon_controller}, at the cost of slightly reducing the platooning performance, compared to a bidirectional topology, while under the effect of a malicious attack, as shown in the simulation results in table~\ref{tab:statistical_comparison_merged} in section~\ref{sec:Simulations}. Extending the method to bidirectional topologies would require a redefinition of the control law and safety filter to account for the additional information flow from following vehicles. This is left as a direction for future work.
\end{assumption}

\begin{assumption}[\textit{Motion constraints}]
\label{as:assumption3}
We consider double integrator dynamics and same actuator limits for each vehicle, i.e. a homogeneous platoon that can be controlled in acceleration. The homogeneity assumption is not a strict requirement, yet we keep it for the sake of clarity, since it simplifies the analysis presented in section~\ref{sec:platoon_controller}. We also neglect external forces such as aerodynamic drag and rolling friction, since we assume an additional control input can always be added on top of our controller's output to compensate for them, as in~\cite{zheng2016distributed}. Furthermore, even simpler models that use velocity control, have been shown to be sufficient for experimental validation on robotic platforms similar to ours~\cite{experimental_validation}. While considering the vehicle dynamics as a second order system simplifies the analysis and is common in platooning literature, we acknowledge that real actuators cannot instantaneously change acceleration. In our experimental implementation, we address this by using a low-level acceleration tracking controller (see Appendix~\ref{sec:appendix}), which ensures smooth and realistic actuation. 
 \end{assumption}
 
\begin{assumption}[\textit{Attacker model}]  \label{ass:4}
The attacker can act both at the platooning control level and at the platooning coordinator level. In the former case it can manipulate the information regarding the vehicle's intended motion that is sent to its follower ,i.e. its acceleration. In the latter, the attacker can communicate a fictitious platooning configuration. A compromised vehicle is one whose outbound communication is manipulated by a malicious agent, while its inbound communication remains functional. This assumption is motivated by the possibility that an attacker has breached authentication protocols for a given vehicle, see Section~\ref{sec:attack feasibility}. Notice that multiple vehicles may be attacked at the same time.
\end{assumption}

\begin{assumption}[\textit{Platoon velocity and spacing}]
We assume that each vehicle in the platoon knows the desired platoon velocity $v^D$ and the desired inter-vehicle distance $d$. Moreover, we constrain the vehicles' velocity to positive values and not to exceed a given maximum value.
\end{assumption}

\subsection{Attack feasibility and motivations}\label{sec:attack feasibility}

Concerning the vulnerability of the on-board sensors, global positioning information is susceptible to spoofing attacks on GNSS technologies, this is especially critical for aerospace applications~\cite{GPS_spoofing_uavs}. However, for automotive applications we consider GNSS spoofing not to be safety critical, since collision avoidance will depend on local positioning information provided by sensors such as RADAR, cameras and LIDAR. Concerning the latter, recent works have shown that it is possible to compromise the LIDAR by removing obstacles or pedestrians~\cite{lidar_spoofing}, however this requires a physical device in close proximity to the target vehicle, making such attacks difficult in practice. Therefore, throughout the remainder of this paper, we assume that a potential malicious attacker does not have enough resources to tamper with the vehicle's state measurements. 

On the other hand, data transfer within a platoon is usually managed through a Dedicated Short Range Communications (DSRC) in a Vehicular Ad-hoc Network (VANET).
According to~\cite{amoozadeh2015security}, various types of attacks can be critical to the safety and string stability of platoon formations. Our solution specifically focuses on mitigating attacks at the application layer, in particular, the message manipulation attack (or false data injection).
To address these concerns, state-of-the-art security architectures leverage robust cryptographic systems to effectively counter application layer attacks, such as digital signatures and one-time pads in messages.
Despite these security methods, practical challenges persist in deploying, implementing, and standardizing such security architectures in VANETs. Moreover, when the threat comes from a trusted insider, like a compromised vehicle with a valid certificate, the problem becomes considerably more challenging.
In the following, we propose an attack-resilient platooning, where we assume that the attacker can modify the transmitted information, by exploiting a lack in the security architecture or a valid certificate that allows to override messages.

\section{Attack-Resilient Platooning}
\label{sec:Resilient_platooning}
This section introduces the design of the control and coordination layers of our attack-resilient platooning strategy as well as the tailored attack detection policy.

    \subsection{Method Overview}
\begin{figure}[t]
 \centering
\includegraphics[width=\columnwidth]{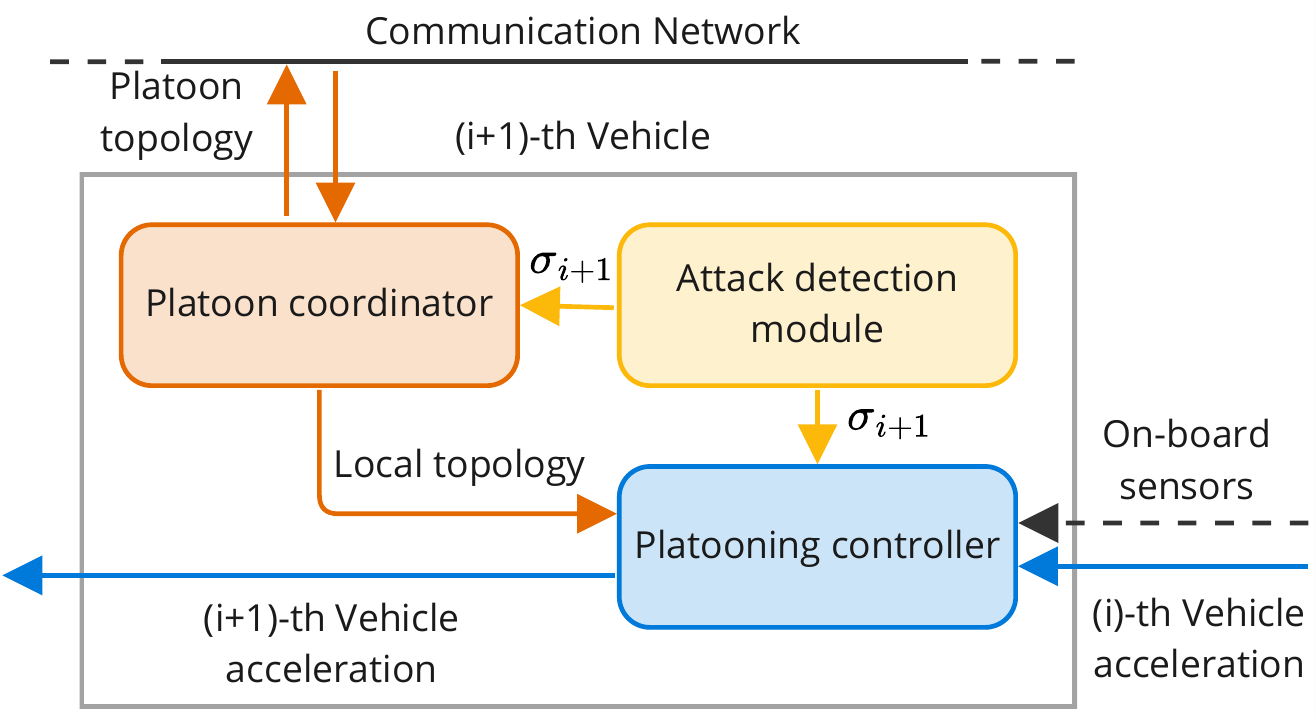}
    \caption{Overview of the proposed distributed platooning scheme.}
  \label{fig:network}
\end{figure}

    Figure~\ref{fig:network}  depicts the overall algorithm scheme for the $(i+1)$-th Vehicle. Each vehicle has its own local platoon coordinator (detailed in Sec.~\ref{sec:platoon_coordinator}) and platoon on-board controller (detailed in Sec.~\ref{sec:platoon_controller}). The coordinator decides in what order to place the vehicles, as well as managing adding or removing vehicles from the platoon. The controller is responsible for keeping the desired distance from the preceding vehicle. More in detail, the coordinator provides a valid local platooning topology to the controller, that is, it indicates which vehicle to follow. The $(i+1)$-th Vehicle's controller synthesizes its acceleration, relying on its on-board sensors (ACC mode) and on the acceleration data received from its preceding vehicle (CACC mode). Figure~\ref{fig:network} also depicts the attack detection module, which provides the controller and the coordinator with $\sigma_{i+1}$, a parameter that measures the communication reliability associated with the preceding vehicle. In the platoon controller $\sigma_{i+1}$ is used to switch off CACC mode and revert back to ACC mode if an attack is detected. In the coordinator $\sigma_{i+1}$ is used to trigger a topology change in order to isolate a single compromised vehicle. We motivate this design choice with the objective of maximizing platooning performance according to the available information concerning the presence of attacks, while still providing safety guarantees. Indeed during nominal conditions CACC ensures best performance in terms of keeping a fixed inter-vehicle distance, as shown in Table~\ref{tab:distances_sin_v}. When an attack is present but still undetected we ensure collision avoidance guarantees thanks to the safety filter that limits the authority of the feed-forward term in the CACC. Once an attack has been detected we switch off the communication between the compromised vehicle and it's follower. We do this because even if the attack is unable to cause a collision, it may still increase the inter-vehicle distance and consequently fuel consumption. This is shown in Figure~\ref{fig:exp2}. This strategy applies to each vehicle and therefore is a valid mitigation approach even if all communication channels between vehicles are compromised, as shown in section~\ref{sec:Simulations} table~\ref{tab:statistical_comparison_merged}. If, however, only one vehicle is affected by an attack, since operating in ACC mode is suboptimal, we then rely on the coordinator to position the compromised vehicle at the end of the platoon. In this way it has no followers, and thus all vehicles resume nominal CACC operation. Notice that we assume only the communication to be affected by an attack, which means that the last vehicle in the platoon, despite having corrupted outbound communication, is still able to receive inbound data from its predecessor and operate normally.

\subsection{Local platooning controller}
\label{sec:platoon_controller}
\begin{figure}[t]
 \centering
\includegraphics[width=1\columnwidth]{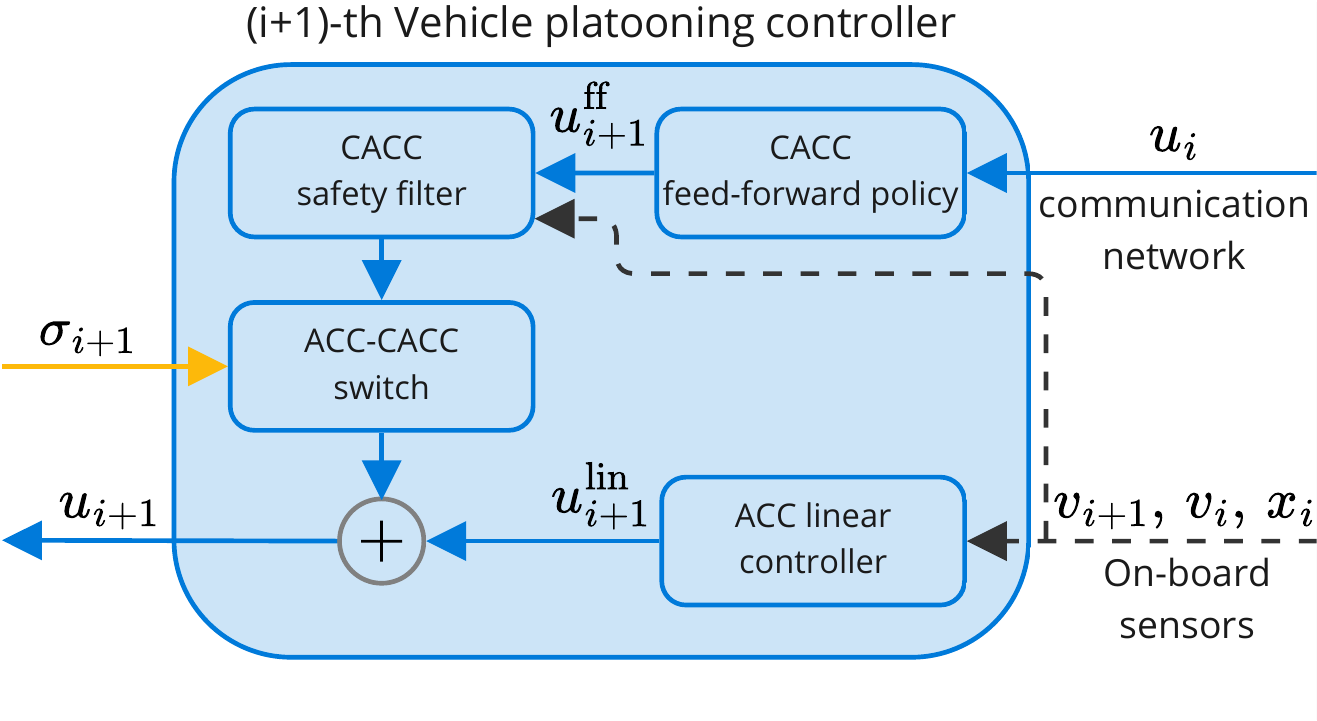}
    \caption{Overview of the proposed low-level platooning controller. The CACC feed-forward policy is detailed in equation~\eqref{eq:uff_policy_generic}, the CACC safety filter corresponds to equation~\eqref{eq: uff}, the ACC-CACC switch refers to equation~\eqref{eq:ACC-CACC switch} and the ACC controller is described in equation~\eqref{eq:control}.}
  \label{fig:platooning controller}
\end{figure}

We build our platooning controller starting from the ACC module using a modified version of the string-stable longitudinal linear controller proposed in~\cite{de1999stability}.  We design a gain tuning procedure that explicitly takes actuator limits into account and is able to provide both string stability and collision avoidance guarantees. We then add a feed-forward predictive term that is essential to improve platooning performance during nominal conditions. We then consider potential false data injection attacks and add a safety filter consisting of control authority constraints to the feed-forward policy. We finally add an ACC-CACC switch to disable the feed-forward action if the communication with the preceding vehicle is deemed unreliable.
Figure~\ref{fig:platooning controller} shows an overview of the platoon controller. In deriving the safety conditions (e.g., Eq.~\ref{eq:p_max complicated}), we assume constant acceleration during braking phases and neglect higher-order dynamics such as actuator lag or jerk as detailed in \textit{assumption 3} in section~\ref{sec:Background}. These simplifications allow us to derive interpretable and implementable safety constraints while maintaining robustness in practice, as validated by our simulations and experiments.

\subsubsection*{ACC Linear controller}
In the following section we provide a linear controller to achieve ACC mode platooning, that given the desired platooning velocity $v^D$ and inter-vehicle distance $d$, takes the vehicle actuation limits into account and provides the range of controller parameters such that string stability and collision avoidance are guaranteed. The linear controller is a modified version of the control law reported in~\cite{de1999stability}[Sec. 4.3].
Let us indicate with $x_{i+1}=\begin{bmatrix} p_{i+1} & v_{i+1} \end{bmatrix}^\top$ the state of the $(i+1)$-th vehicle, where  $p_{i+1},v_{i+1}$ are the $(i+1)$-th vehicle's position and velocity, respectively. The state of the preceding vehicle is $x_{i}$. According to Assumption~\ref{as:assumption3}, the vehicles can be controlled by acceleration commands\footnote{We also assume that a reliable vehicle model, which can convert the required acceleration into the vehicle inputs, is available.} and the vehicle actuation limits are known, namely the maximum acceleration $u_{\max}>0$, maximum braking $u_{\min}<0$ and maximum vehicle velocity $v_{\max}$. The acceleration-controlled vehicles are represented by the following linear dynamics:
\begin{equation}\label{eq:system}
    \dot{x}_{i+1}(t)\!=\! Ax_{i+1}(t)\! +\! B u^{\text{lin}}_{i+1}(t) , \forall i = 1, \dots ,N-1, t\geq 0
\end{equation}
where 
$A = \begin{bmatrix}
    0 &1\\0&0
\end{bmatrix},$
$B=\begin{bmatrix}
    0\\ 1
\end{bmatrix}$. The control action is: 
\begin{equation}\label{eq:control}\begin{split}
        u^{\text{lin}}_{i+1} = &-k(p_{i+1}-p_{i}+d)-kh(v_{i+1}-v^D) \\ &- c (v_{i+1}-v_{i}), 
\end{split}
\end{equation}
where $k, c, h > 0$ are the tuning parameters.
To enforce string stability we  first derive an expression of the transfer function associated to equation~\eqref{eq:system}. Since the objective of the platoon is to travel at a desired speed $v^D$ we define the time-dependent reference position for vehicle $i$ as $p_i^{\text{ref}}=p_i(t=0)+v^Dt$ and since vehicle $(i+1)$ should follow vehicle $i$ at a distance of $d$ we define $p_{i+1}^{\text{ref}}=p_i(t=0)+v^Dt-d$. We then define $\hat{p}_{i}=p_i-p_i^{\text{ref}}$ and $\hat{p}_{i+1}=p_{i+1}-p_{i+1}^{\text{ref}}$. In equation~\eqref{eq:control} we rewrite $u^{\text{lin}}_{i+1}$ as $u^{\text{lin}}_{i+1}=\ddot{p}_{i+1}$ and substitute the expressions for $p_i=\hat{p}_{i}+p_i^{\text{ref}}$, $p_{i+1}=\hat{p}_{i+1}+p_{i+1}^{\text{ref}}$ and their derivatives to obtain the expression of the transfer function $G_{i+1}(s)$:
\begin{equation}\label{eq:transfer function G}
    G_{i+1}(s) = \frac{\hat{p}_{i+1}(s)}{\hat{p}_{i}(s)} =\frac{c s + k}{s^2+(c+hk) s + k}.
\end{equation}
The magnitude of the transfer function $G_{i+1}(s)$ maps the scale factor between the deviation from the reference position---and consequently also reference velocity $v^D$ of the $(i+1)$-th vehicle with respect to its preceding vehicle. Notice that $G_{i+1}(s)$ also applies to relative quantities:
\begin{equation}\label{eq:relative transfer function G}
    \frac{p_{i+2}-p_{i+1}+d}{p_{i+1}-p_{i}+d}=\frac{\hat{p}_{i+2}-\hat{p}_{i+1}}{\hat{p}_{i+1}-\hat{p}_{i}} =\frac{\hat{p}_{i+1}(G-1)}{\hat{p}_{i}(G-1)}=\frac{\hat{p}_{i+1}}{\hat{p}_{i}} ,
\end{equation}
where we dropped the $(s)$ dependency to compact the notation.
According to the definition of string stability, we require that $\|G_{i+1}(j\omega)\|<1$, $\forall$ $\omega >0$, $\forall i=1,\dots,N-1$. This can be achieved by ensuring that at least one pole is smaller than the zero. Additionally, we further require that the controlled system is not underdamped, this avoids the follower's position overshooting it's reference during transients and leading to collisions. These two requirements can be written as:
\begin{align}\label{eq: string stab constraint}
    \frac{1}{2}(c+hk)-\frac{1}{2}\sqrt{(c+hk)^2-4k}<\frac{k}{c}\\\label{eq: crit damp constraint}
    (c+hk)^2-4k>0.
\end{align}

\begin{figure}[t]
 \centering
\includegraphics[width=1\columnwidth]{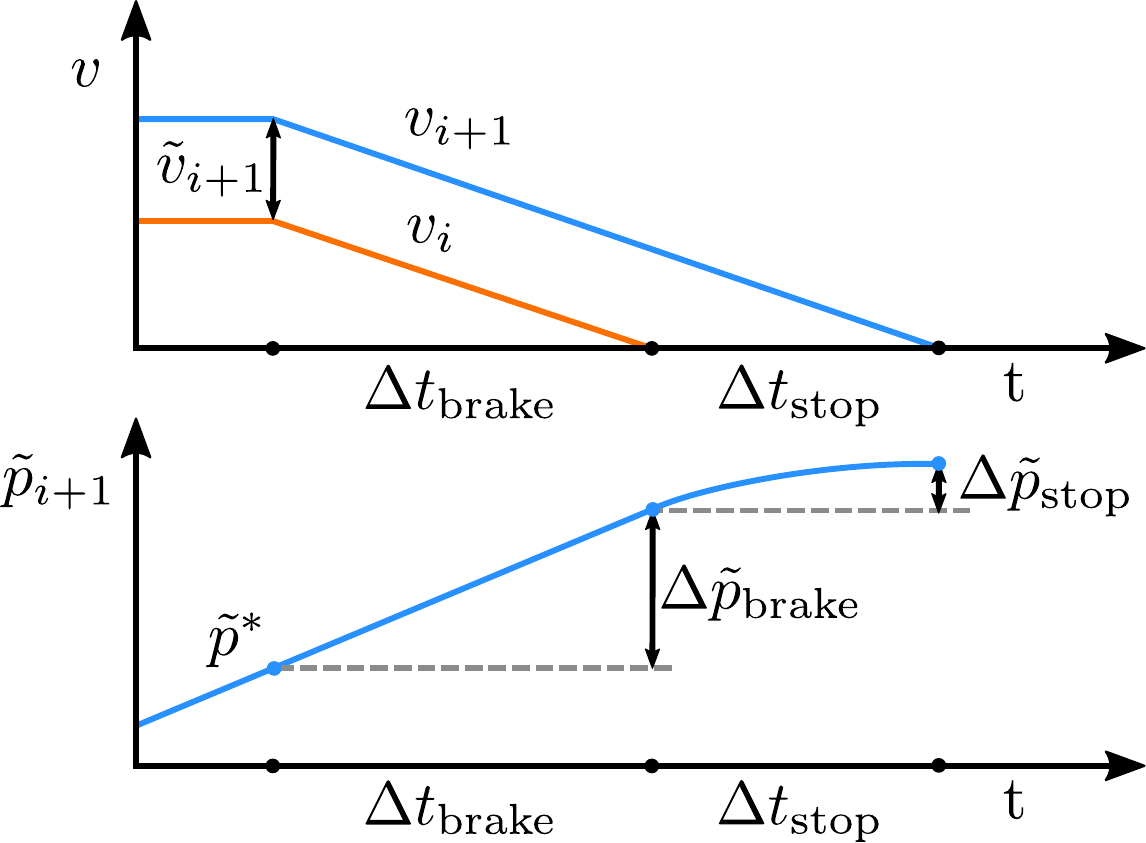}
    \caption{Maximum reachable $\Tilde{p}$ during an emergency brake.}
  \label{fig: reacheble p analysis}
\end{figure}

Tuning the controller to ensure string stability does not yet guarantee absence of inter-vehicle collisions when the actuators are bounded. We thus need to derive additional conditions on the tuning parameters from the collision avoidance requirements. Let us introduce new state variables $\Tilde{v}_{i+1}=v_{i+1}-v_i$ and $\Tilde{p}_{i+1}=p_{i+1}-p_i+d$. By defining $\Tilde{x} = \begin{bmatrix} \Tilde{p} & \Tilde{v} \end{bmatrix}^\textrm{T}$, we can subtract the absolute dynamics~\eqref{eq:system} of the $i$-th vehicle from the $(i+1)$-th vehicle and write the  system's relative dynamics as: \begin{equation}\label{eq:system_relative_1}
    \dot{\Tilde{x}}_{i+1}(t) = A\Tilde{x}_{i+1}(t) + B (u_{i+1}(t) -u_i(t)) , t\geq 0. 
\end{equation}
Note that we also added $d$ to either side of the first row in~\eqref{eq:system_relative_1}. Under the effect of actuator saturation, without explicitly considering braking time, the reachable set of $\Tilde{p}_{i+1}$ becomes unbounded. This can be seen from equation~\eqref{eq:system_relative_1}. 
If $u_{i+1} = u_{i} = u_{\min}$, that is, when both vehicles exert their maximum braking capabilities, $\dot{\Tilde{v}}_{i+1} = 0$. If $\Tilde{v}_{i+1}>0$ (which is the case for a string stable controller) then $\Tilde{p}_{i+1}$ will keep growing until a collision occurs, that is, $\Tilde{p}_{i+1}>d$. We now provide an expression for the maximum reachable $\Tilde{p}_{i+1}$, and provide conditions on $k,c,h$ such that $\Tilde{p}_{i+1}\leq d \,\, \forall\,\, t\in[0,\infty)$.
The controller braking saturation line in the $\Tilde{v}$-$\Tilde{p}$ plane is
\begin{align*}
    -k \Tilde{p} - c\Tilde{v}-kh(v-v^D)=u_{\min}.
\end{align*}
We chose $k$ such that the braking saturation line stays always below the point $(0,d)\,\,\forall\,\,v\in[0,v_{\max}]$, this ensures that for $\Tilde{v}\geq0$, braking saturation will always occur before a potential collision. This can be written as:
\begin{align}
    k = \frac{-u_{\min}}{d-h v^D}.
\end{align}
Let us consider an emergency braking scenario where the preceding vehicle suddenly brakes by applying $u_{i}=u_{\min}$. We define the maximum reachable $\Tilde{p}$ as
\begin{align}\label{eq:p_max simple}
    \Tilde{p}_{\max}=\Tilde{p}^* + \Delta\Tilde{p}_{\text{brake}}+ \Delta\Tilde{p}_{\text{stop}}
\end{align}
Where $\Tilde{p}^*$ is the initial value of $\Tilde{p}$ when the $(i+1)$-th vehicle reaches braking saturation, $\Delta\Tilde{p}_{\text{brake}}$ is the distance travelled while both vehicles are braking and $\Delta\Tilde{p}_{\text{stop}}$ is the distance travelled by the $i+1$-th vehicle once the preceding vehicle has already stopped. In Figure~\ref{fig: reacheble p analysis} we provide a visualization of the absolute velocity $v$ and relative position $\Tilde{p}$ during an emergency brake.

We define $\Tilde{p}^*$ as:
\begin{align}\label{eq: p*}
    \Tilde{p}^*=d-\frac{c}{k}\Tilde{v}.
\end{align}
Note that this is a conservative estimate since saturation will actually occur for lower values of $\Tilde{p}$ for $v>0$. $\Delta\Tilde{p}_{\text{brake}}$ is defined as $\Delta\Tilde{p}_{\text{brake}}=\Tilde{v}\Delta t_{\text{brake}}$, where $\Delta t_{\text{brake}}$ is the time it takes the preceding vehicle to stop, since $v_{i+1}=v_{i}-\Tilde{v}$ we can write $\Delta\Tilde{p}_{\text{brake}}$ as:
\begin{align}\label{eq: Dp brake}
    \Delta\Tilde{p}_{\text{brake}}=\Tilde{v}\frac{v_{i+1}-\Tilde{v}}{-u_{\min}}.
\end{align}
Once the preceding vehicle has stopped, the $i+1$-th vehicle has velocity $v_{i+1}=\Tilde{v}$, thus $\Delta\Tilde{p}_{\text{stop}}$ is defined as:
\begin{align}\label{eq: Dp stop}
    \Delta\Tilde{p}_{\text{stop}}=\frac{1}{2}\frac{\Tilde{v}^2}{-u_{\min}}
\end{align}
By substituting equations~\eqref{eq: p*},~\eqref{eq: Dp brake} and~\eqref{eq: Dp stop} in equation~\eqref{eq:p_max simple} we can write $\Tilde{p}_{\max}$ as:
\begin{align}
    \Tilde{p}_{\max}=d-\left( \frac{c}{k}-\frac{v}{-u_{\min}}\right)\Tilde{v}-\frac{1}{2}\frac{\Tilde{v}^2}{-u_{\min}}
\label{eq:p_max complicated}
\end{align}

\begin{figure}
 \centering
\includegraphics[width=\columnwidth]{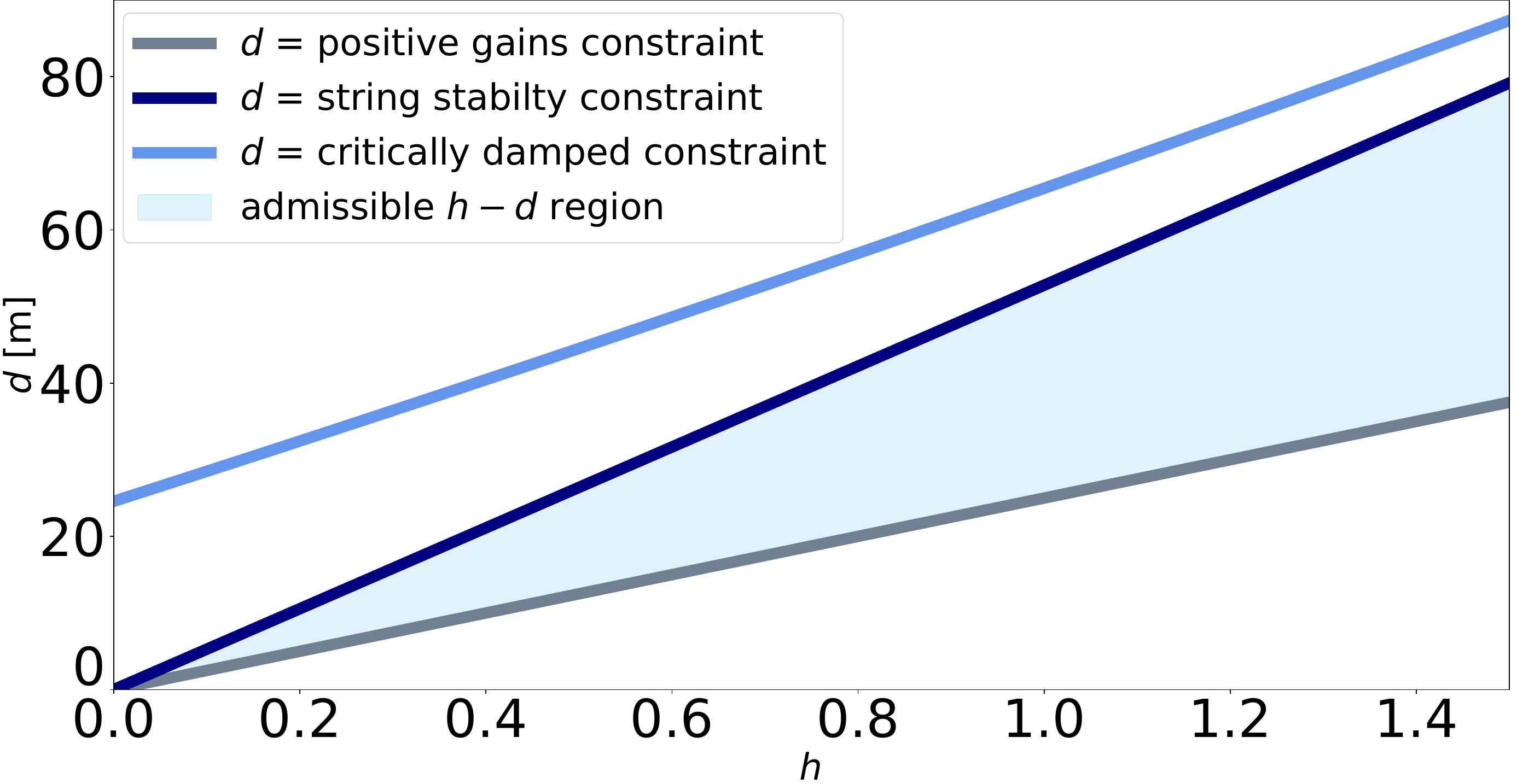}
    \caption{Gain tuning graph showing the combinations of $d$ and $h$ that guarantee string stability and collision avoidance for vehicle parameters $v_{\max} = 100$ (km/h), $u_{\min} = -0.8 g$, $u_{\max} = 0.5 g$, and platooning velocity $v^D=90$ (km/h).}
  \label{fig: gain tuning graph}
\end{figure}

Notice that $\Tilde{p}_{\max}$ is monotonically increasing with respect to $v$, so by requiring that $\Tilde{p}_{\max}\leq d$ for $v=v_{\max}$, we guarantee collision avoidance. This can be achieved by choosing $c$ as:
\begin{align}
    c = \frac{v_{\max}}{d-hv^D}.
\end{align}
We now have expressions for $k$ and $c$ as a function of $d$ and $h$, so to conclude the tuning procedure we represent conditions~\eqref{eq: string stab constraint},~\eqref{eq: crit damp constraint} and $k,c,h>0$ in the $h$-$d$ plane. By choosing a valid combination of $h$ and $d$ we can guarantee string stability and collision avoidance while accounting for saturation constraints. Figure \ref{fig: gain tuning graph} shows the possible valid $h$-$d$ combinations for vehicle parameters $v_{\max} = 100$ (km/h), $u_{\min} = -0.8 g$, $u_{\max} = 0.5 g$, and platooning velocity $v^D=90$ (km/h). 
It is worthwhile mentioning that for a given $d$ the closer $h$ is to the $k,c>0$ line, the higher the gains will be, for $h\rightarrow\frac{d}{v^D}$ $k,c\rightarrow\infty$. Yet it is still possible to chose very small values of $d$, in the range $5-10$m such to provide high aerodynamic and fuel efficiency benefits to the platoon.

\subsubsection*{Feed-forward policy and safety filter}
To enhance performance, we rely on V$2$V communication to achieve CACC platooning. We introduce a communication-based feed-forward term $u^{\text{ff}}$ and define the $(i+1)$-th vehicle's acceleration as $u_{i+1}=u^{\text{lin}}_{i+1}(t)+u_{i+1}^{\text{ff}}(t)$. We can now rewrite Eq.~\eqref{eq:system_relative_1} as follows:
\begin{equation}\label{eq:system_relative}
    \dot{\Tilde{x}}_{i+1}(t) = A\Tilde{x}_{i+1}(t) + B (u^{\text{lin}}_{i+1}(t)+u_{i+1}^{\text{ff}}(t) -u_i(t)),
\end{equation}


The relevance of $u_{i+1}^{\text{ff}}$ can be clearly appreciated by viewing the preceding vehicle's acceleration $u_i$ as a disturbance, indeed if a reliable measure of the latter is available, we can design $u_{i+1}^{\text{ff}}$ to compensate for it and even improve the platooning performance. In this section, to derive the safety conditions, we need to make no assumptions on the specific choice of feed-forward policy, it could be based on MPC, sliding mode controller~\cite{drakunov1992sliding}, or even a simpler policy such as $u_{i+1}^{\text{ff}}(t) = u_{i}(t)$. We refer to the generic feed-forward policy as:
\begin{align}\label{eq:uff_policy_generic}
    u_{i+1}^{\text{ff}}(t) = \pi(u_i(t)).
\end{align}

Note that in~\eqref{eq:uff_policy_generic} $\pi$ may also depend on the $i$-th and $(i+1)$-th vehicles states, yet we highlight the dependency on $u_i(t)$ because this data is strictly necessary for any kind of feed-forward policy and needs to be shared over the communication network. A well designed policy should improve the overall performance of the platoon, yet if the $(i+1)$-th vehicle receives compromised $u_i$ acceleration data, a malicious entity could exploit $u_{i+1}^{\text{ff}}$ to take control of the $(i+1)$-th vehicle and potentially lead it to unsafe states.

To guarantee collision avoidance under corrupted $u_i$ acceleration data we implement two additional conditions on $u_{i+1}^{\text{ff}}$, leading to the following \emph{safety filter} definition: 
\begin{subnumcases}{u_{i+1}^{\text{ff}}=\label{eq: uff}}
  0 & if $\Tilde{p}_{i+1}\geq d-\frac{c}{k}\Tilde{v}_{i+1}$ \label{eq: uff_no_sat_lin} \\
  \hat{u}^{\text{ff}}_{\max}(v_{i+1}) & if $\pi(u_i)\geq u^{\text{ff}}_{\max}(v_{i+1})$\label{eq: uff_max_val}\\
  \pi(u_i) & otherwise \label{eq: uff otherwise} 
\end{subnumcases}

\begin{figure}[t]
 \centering
\includegraphics[width=0.9\columnwidth]{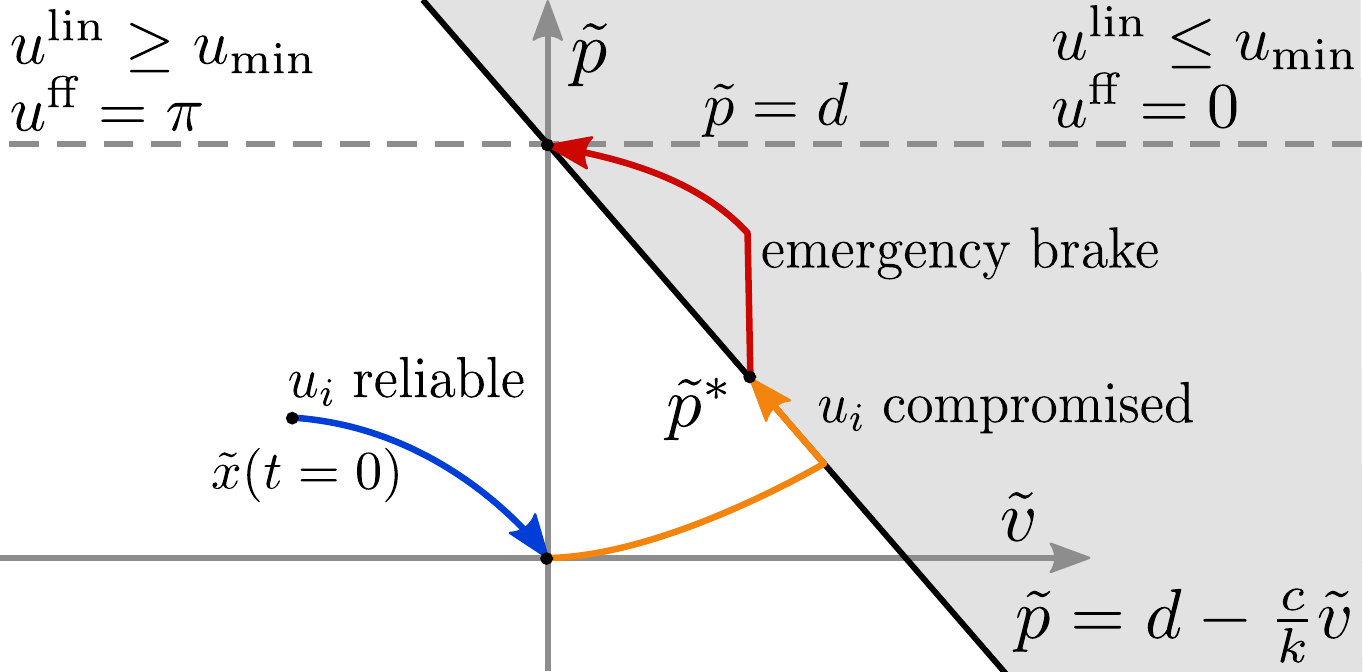}
    \caption{State trajectory example in the $\Tilde{p}$-$\Tilde{v}$ plane. From an initial condition $\Tilde{x}(t=0)$ the system converges to the origin while a reliable measure of $u_i$ is available (blue arrow). Subsequently a malicious attack communicates false acceleration data and induces vehicle $(i+1)$ to approach its predecessor, yet no collision occurs thanks to additional constraints that limit the control authority of $u^{\text{ff}}_{i+1}$ (orange arrow). Finally, the leading vehicle performs an emergency brake (red arrow) but collision is avoided thanks to the linear controller design process that accounts for actuation saturation (shaded area).}
  \label{fig:relative-pv plane}
\end{figure}

Where we define $u^{\text{ff}}_{\max}(v_{i+1}) =  k(\alpha d+h(v_{i+1}-v^D))$, $\alpha \in [0,1]$. Condition~\eqref{eq: uff_no_sat_lin} essentially triggers an emergency braking maneuver and ensures that the collision avoidance guarantees provided by the ACC linear controller are still verified even in the presence of a compromised $u_i$ measure. This is because the feed-forward action is disabled if the current state in the $(\Tilde{v},\Tilde{p})$ plane is above the linear controller braking saturation line for $v_{i+1}=0$. In this case $u_{i+1}=u^{\text{ff}}_{i+1}+u^{\text{lin}}_{i+1}=u_{\min}$.
Condition~\eqref{eq: uff_max_val} is added to limit the number of times the emergency braking condition~\eqref{eq: uff_no_sat_lin} is triggered. By setting the maximum value of $u^{\text{ff}}_{i+1}$ to $k(\alpha d+h(v-v^D))$, we ensure that $u^{\text{lin}}_{i+1}+u^{\text{ff}}_{i+1}\leq 0\,\,\, \forall \,\, \Tilde{p}  \geq \alpha d-\frac{c}{k} \Tilde{v}$, i.e. under a compromised $u^{\text{ff}}_{i+1}$, the state of vehicle $(i+1)$ will remain below the line described in equation~\eqref{eq: p*} and condition~\eqref{eq: uff_no_sat_lin} will not be activated.  This means that in order to trigger an emergency maneuver, an attacker needs to compromise both the vehicle $(i+1)$, inducing it to accelerate, and the vehicle $i$, inducing it to brake. Values of $\alpha<1$ will introduce an additional safety margin, since the minimum inter-vehicle distance under a compromised $u^{\text{ff}}$ will be $(1-\alpha)d$, at the expense of more stringent limitations on $u^{\text{ff}}$. Note that even if $\alpha=1$ collision avoidance is still guaranteed thanks to condition~\eqref{eq: uff_no_sat_lin}. In Figure~\ref{fig:relative-pv plane} we show an example state trajectory in the $\Tilde{p}$-$\Tilde{v}$ plane for $\alpha = 1$.

\subsection{Attack detection module}
The platooning controller presented in the previous sections is robust to adversarial attacks, in particular, it can provide collision avoidance guarantees even when the communication channel between all vehicles is compromised. However, a malicious entity that carries out an attack at the communication network level could still compromise the performance of the platoon. For example by injecting false acceleration data, it could increase the inter-vehicle distance or induce large oscillations in the velocity. To mitigate such an attack we rely on an \emph{attack detection module}, that runs on each vehicle, constantly evaluates the reliability of $u_{i}$, and outputs a variable $\sigma_{i+1}\in\{0,1\}$, where $\sigma_{i+1} = 1$ if vehicle $i+1$ trusts it's predecessor's $u_{i}$ information and $\sigma_{i+1} = 0$ otherwise. We thus  redefine the feed-forward controller in~\eqref{eq: uff} as:
\begin{equation}\label{eq:ACC-CACC switch}
u_{i+1}^{\text{ff}} = \sigma_{i+1} u_{i+1}^{\text{ff}},
\end{equation}
that is, we disable the feed-forward action $u_{i+1}^{\text{ff}}$ if $\sigma_{i+1} = 0$ in order to protect against attacks aimed at degrading the performance of the platoon. Note that this strategy is intrinsically robust to denial of service attacks, as this is equivalent to setting $\sigma_{i+1}=0$.

 There are multiple ways to estimate the trustworthiness of the data received from the preceding vehicle, such as checking consistency between communication and on-board sensors, or evaluating and recognizing a performance degradation. More sophisticated techniques can be designed based on particle filters~\cite{jahanshahi2018attack}, sliding mode observers~\cite{bissmeyer2012assessment} or set membership estimation~\cite{set_membership_attack}. In the present work we opted to implement an observer based approach due to its simplicity. Additionally, as in~\cite{set_membership_attack} and in the present paper, in vehicular platooning it is typical to assume that a malicious attack is most likely to occur on the communication channels, rather than on the on-board sensors. Since vehicles can rely on the latter to estimate the current state, we exploit this peculiarity to separate the state estimation from the attack detection functionality and design a tailored Kalman filter-based detector module. 

Since each vehicle trusts its own sensors and actuators, the effects of a compromised $u_i$ measure are visible when estimating $\Tilde{v}_{i+1}$. As pointed out in~\cite{stelthiness_kalman_filter_attack}, the filter gain $K$ reaches an equilibrium after a certain number of time steps based solely on the ratio between the process noise matrix $Q$ and the measurement noise $n$, it is thus reasonable to evaluate the equilibrium gain offline and use it for state estimation. A constant gain Kalman filter that estimates $\Tilde{v}_{i+1}$ is represented by following equations:
\begin{align}
    \Tilde{v}_{i+1}^{-}&=\hat{\Tilde{v}}_{i+1}^k+dt(u_{i+1}^k-u_i^k)\label{eq:kalman prediction}\\
    \hat{\Tilde{v}}_{i+1}^{k+1}&=(1-K)\Tilde{v}_{i+1}^{-} + K \Tilde{v}_{i+1}^{k+1},\label{eq:kalman update}
\end{align}

Where $k$ indicates the time instant, $dt$ is the time step, $\Tilde{v}_{i+1}^{-}$ is the expectation,  $\hat{\Tilde{v}}_{i+1}$ is the estimated relative velocity and $\Tilde{v}_{i+1}^{k+1}$ is the measured relative velocity. As discussed in~\cite{stelthiness_kalman_filter_attack}, the effect of a malicious attack that remains undetected is bounded by the Kalman filter properties, i.e., by the value of $K$. In vehicular platooning this is clear by looking at equation~\eqref{eq:kalman prediction}, since $u_i$ is the only quantity that can be affected by a malicious attacker, for $K\rightarrow1$ the estimated $\hat{\Tilde{v}}_{i+1}$ will predominantly be affected by the measured state $\Tilde{v}_{i+1}$. Conversely if $K\rightarrow0$ the predominant contribution to the $\hat{\Tilde{v}}_{i+1}$ update will be based on $\Tilde{v}_{i+1}^{-}$ and the estimated relative velocity will be more sensitive to a malicious attack on $u_i$. On a real system the available sensors are typically sufficiently accurate and provide a reliable measure of $\Tilde{v}_{i+1}$ without the need to run the Kalman filter described in equations~\eqref{eq:kalman prediction} and~\eqref{eq:kalman update}, thus by choosing lower values of $K$ it can be focused towards verifying the trustworthiness of $u_i$, rather than providing a good estimate of $\Tilde{v}_{i+1}$. This is achieved by evaluating the residual $r$, defined as:
\begin{align}
    r_{i+1}^k= |\hat{\Tilde{v}}_{i+1} - \Tilde{v}_{i+1}|.
\end{align}

We then consider the communication channel to be under attack if $r_{i+1}^k$ exceeds a certain threshold value $\Bar{r}$. Setting the values $K$ and $\Bar{r}$ entails striking a compromise between reactiveness and false detections. In particular lower values of $K$ also increase sensitivity to noise on $u_i$, $u_{i+1}$ and $\Tilde{v}_{i+1}$, and $\Bar{r}$ needs to be calibrated in order to avoid missed detections while minimizing false positives.

\subsection{Platoon coordinator}\label{sec:platoon_coordinator}
An additional mechanism to mitigate attacks can be implemented by relying on the platoon coordinator.
 In fact, the platoon coordinator can reorganize the platoon to isolate the effects of a compromised vehicle, i.e. a vehicle whose outbound communication is unreliable. We leverage the predecessor-follower topology and rearrange the platoon such that the compromised vehicle is in the last position, in this way it has no follower to communicate with. In addition it can handle merging and splitting requests. 

We propose a distributed strategy that guarantees the successful execution of the aforementioned operations, even in the presence of a single malicious vehicle that communicates false information at the platoon coordinator level.

Each vehicle, numbered $i$, needs to build a vector $\delta_i $, which includes information about the preceding and the following vehicles, and broadcasts it to the network.
The broadcasting operation can be done in a distributed fashion by relying on existing methods~\cite{tonguz2007broadcasting}.

The entries $\delta_i = \begin{bmatrix}
    \text{id}_{i,i-1} & \text{id}_{i,i+1}
\end{bmatrix}$ correspond to a unique number associated with the preceding and the following vehicle (no preceding or no following vehicle is indicated with $\text{id}_{i,i-1} =0$ and  $\text{id}_{i,i+1}=0$, respectively).
Additionally, when the trustworthiness value, provided by the estimator module, becomes $\sigma_{i} = 0$, to indicate a severed communication link, we set $\text{id}_{i,i-1} =0$.

We define \emph{correct platooning conditions} when: \emph{1.} there is only one vehicle without the preceding vehicle (the leader) and only one without the following vehicle (the last vehicle), \emph{2.} the matrix $\mathcal{D}= [\delta_1, \delta_2, \dots, \delta_N]^\textrm{T}$ has to describe a connected topology and finally, \emph{3.} it has to be consistent, that is, if vehicle A is the preceding vehicle of B, then B is the following vehicle of A.

If the \emph{correct platooning conditions}  are violated, the platoon coordinator comes into play.
We inspect three cases of interest that require the intervention of the platooning coordinator:
\emph{(i)} \emph{reorganization to isolate a compromised vehicle}: the $i$-th vehicle receives unreliable acceleration data, and sets $\text{id}_{i,i-1}=0$, the correct platooning conditions are violated, since there are two vehicles with $\delta_i = \begin{bmatrix}
        0 & \text{id}_{i,i+1} 
    \end{bmatrix}$.
    \emph{(ii)} \emph{merging request}: a vehicle wants to join the platoon. In this case the vehicle sends a request to the network by communicating its vector $\delta_i = \begin{bmatrix}
        0 & 0 
    \end{bmatrix}$ and a new row is added to the   $\mathcal{D}$ matrix;
        \emph{(iii)} \emph{splitting request}: a vehicle wants to leave the platoon. The vehicle decides to not communicate with the network anymore and its row is removed from the $\mathcal{D}$ matrix. The preceding and the following vehicle of the departing agent remain without a following and a preceding vehicle, respectively. This leads to a violation of the conditions on  $\mathcal{D}$, that is, two vehicles with $\delta_i = \begin{bmatrix}
        0 & \text{id}_{i,i+1} 
    \end{bmatrix}$ and two vehicles with $\delta_i = \begin{bmatrix}
         \text{id}_{i,i-1}  & 0 
    \end{bmatrix}$.

We assumed that each vehicle in the system has a copy of the matrix  $\mathcal{D}$. The platooning coordinator on each vehicle, constantly checks the correctness of the  $\mathcal{D}$ matrix, that is, the correctness of the platooning conditions. When one or more conditions are violated each vehicle solves the following optimization problem:
\begin{equation}\label{eq:optimization1}
\begin{split}
    &\argmax_{ \mathcal{D}^\star} \sum_{i=1}^N f(\text{id}_{i,i-1},\text{id}_{i,i-1}^{\star}) +f(\text{id}_{i,i+1},\text{id}_{i,i+1}^{\star})  \\
    &\text{s.t. \emph{correct platooning conditions},}
\end{split}
\end{equation}
where $f(a,b)=1$, if $a=b$, $f(a,b)=0$, if $a\neq b$. Notice that the optimization cost function weights the difference between the old configuration  $\mathcal{D}$, which violates the \emph{correct platooning conditions}, and the new configuration  $\mathcal{D}^\star$. This means that the new solutions will differ from the previous graph topology as little as possible to ensure \emph{correct platooning conditions}. 
The algorithm scalability in general may be an issue, however, for our application there are no stringent computation time constraints, since the platooning coordinator does not affect safety nor stability. 

\begin{figure}[t]
  \centering
 \includegraphics[width=1\columnwidth]{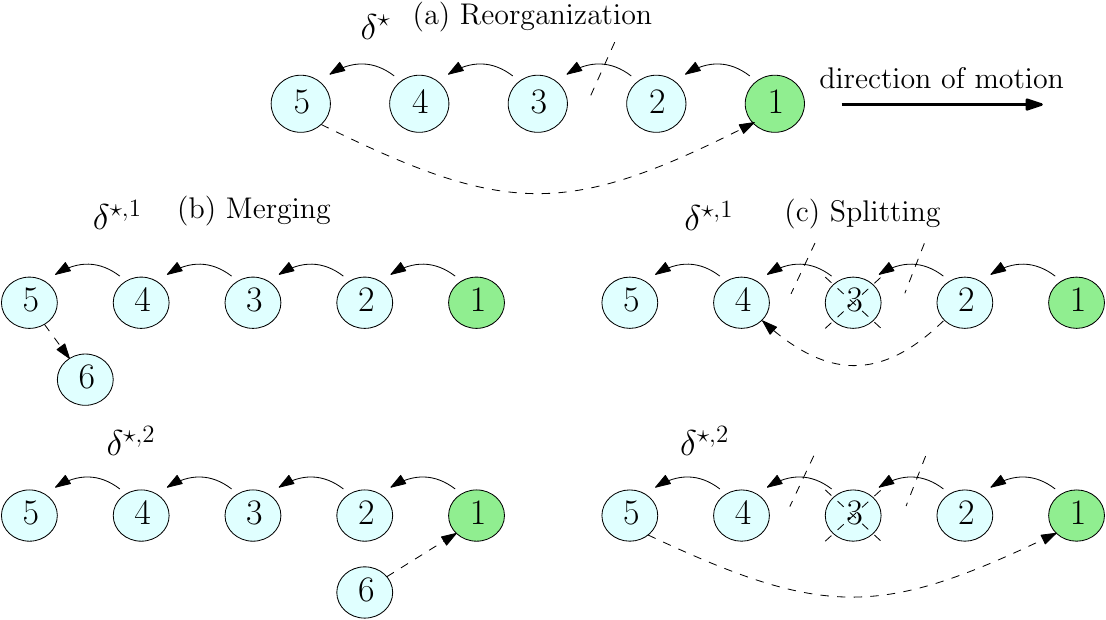}
    \caption{Three example of the platoon coordinator intervention. In (a) we depicted the platoon reorganization, in (b) the merging case, in (c) we represented the splitting case. The dashed arrows are obtained by solving~\eqref{eq:optimization1}.}
  \label{fig:platoon2}
\end{figure}

\begin{example}
Figure~\ref{fig:platoon2} depicts the possible solutions of~\eqref{eq:optimization1} for the three cases: platoon reorganization, merging request and splitting request.
\subsubsection{Platoon reorganization} 
Let us consider Figure~\ref{fig:platoon2}-a, the case where the trustworthiness value $\sigma_{3} =0$. In this particular condition, the optimal solution of~\eqref{eq:optimization1} would  reattach the removed link, yet since that communication channel is now considered unreliable we remove the latter from the admissible solutions. The solution of~\eqref{eq:optimization1} is thus unique, placing vehicle $2$, which communicates suspiciously, in the last position, where it can no longer influence any other vehicle. Notice that despite the attack on the communication channel has been isolated, the vehicle itself is still part of the platoon, since we assume it is still able to receive inbound communication from it's predecessor and remains cooperative. In this scenario there is no external way to assess whether the attack on the last vehicle has been resolved or not. Nevertheless, the platoon has now been alerted and vehicle $2$ is now aware that its own outbound communication has been compromised. In this situation the compromised vehicle could initiate a full system check or implement other ad hoc strategies, yet we consider this to be out of scope of the present paper.
$$\delta = \begin{bmatrix}
    0 & 2 \\ 1 & 3 \\ 0 & 4 \\ 3 & 5 \\ 4 & 0 
\end{bmatrix},\,\,\, \delta^{\star} = \begin{bmatrix}
    5 & 2 \\ 1 & 0 \\ 0 & 4 \\ 3 & 5 \\ 4 & 1 
\end{bmatrix}$$

\subsubsection{Merging case}
Let us consider Figure~\ref{fig:platoon2}-b, the case where vehicle $6$ wants to join the platoon. By solving~\eqref{eq:optimization1} we get two possible solutions. In fact, since the optimization problem minimizes the changes in the platooning configuration, vehicle $6$ could take the place of the leader or of the last vehicle. 
To resolve the ambiguity, a simple rule could be to not change the platoon leader, hence solution $\text{id}^{\star,1}$ would be preferred.
$$ \delta = \begin{bmatrix}
    0 & 2 \\ 1 & 3 \\ 2 & 4 \\ 3 & 5 \\ 4 & 0 \\ 0 & 0
\end{bmatrix},\,\,\, \delta^{\star,1} = \begin{bmatrix}
    0 & 2 \\ 1 & 3 \\ 2 & 4 \\ 3 & 5 \\ 4 & 6 \\ 5 & 0
\end{bmatrix}, \,\,\, \delta^{\star,2} = \begin{bmatrix}
    6 & 2 \\ 1 & 3 \\ 2 & 4 \\ 3 & 5 \\ 4 & 0 \\ 0 & 1
\end{bmatrix} $$

\subsubsection{Splitting case} 

Let us consider Figure~\ref{fig:platoon2}-c, the case where vehicle $3$ wants to leave the platoon. Also in this case, by solving~\eqref{eq:optimization1}, we get two possible solutions, one is simply to link the gap left by vehicle $3$, the other solution instead, is to promote vehicle $4$ as the leader and move vehicle $2$ to the end of the platoon. Similarly to the merging case the  $\delta^{\star,1}$ solution would be preferred, since the platoon leader should not change.
$$\delta = \begin{bmatrix}
    0 & 2 \\ 1 & 0 \\ 0 & 5 \\ 4 & 0 
\end{bmatrix},\,\,\, \delta^{\star,1} = \begin{bmatrix}
    0 & 2 \\ 1 & 4 \\ 2 & 5 \\ 4 & 0
\end{bmatrix},\,\,\, \delta^{\star,2} = \begin{bmatrix}
    5 & 2 \\ 1 & 0 \\ 0 & 5 \\ 4 & 1 
\end{bmatrix}$$
$\hfill\blacksquare$
\end{example}


\subsubsection*{Security issues}

Until now we did not discuss how the platoon coordinator can manage the case where a vehicle in the network broadcasts false topology information. Let us consider the case where the $i$-th vehicle is compromised and communicates a false $\delta_i$. This kind of attack is easily detectable in our proposed framework. In fact, the platoon coordinator constantly checks the correctness of the  $\mathcal{D}$ matrix, and in particular its consistency, as we described before (Property \emph{3.}). In case of inconsistencies it means that a $\delta_i$ message has been manipulated or a trustworthiness value $\sigma_i$ dropped to $0$.
If an attacker communicates a false configuration to the platoon, assuming $N>3$, we have to distinguish two cases: \emph{1)} If even a single entry of $\delta_i$ is changed with another id, there are at least other two vehicles that are in contrast with the information provided by the attacker. By trusting the majority we can identify the attacker and override its messages; \emph{2)} A subtler attack is to communicate to the neighbours to not have a preceding vehicle, emulating a drop in the preceding vehicle's trustworthiness. Since the $\sigma$ values are not shared between the vehicles, it is not trivial to recognize it as an attack. However, we argue that this is not a significant threat, in fact it would involve a single rearrangement of the platoon, after that, the attacker would no longer have a chance to rearrange the platoon any further, since it would become the new leader of the platoon and thus would not have a predecessor in any case.

\section{Simulation results}
\label{sec:Simulations}
In this section we compare our low-level controller against the MPC-based platooning controller~\cite{zheng2016distributed} and the reachability-based robust control approach~\cite{kafash2018constraining}. Notice that for these simulations we don't include any attack-detection components in order to highlight the safety guarantees of the CACC-ACC controller even if an attack remains undetected.

The vehicle parameters are set as in figure~\ref{fig: gain tuning graph}, i.e. actuator limits are $v_{\max}=100$km/h, $u_{\min}=-0.8g$, $u_{\max}=0.5g$. All controllers try to keep an inter-vehicle distance of $d=6$m while traveling at a target velocity of $v_D=90$km/h. For our controller we select the lowest value of $h$ compatible with these parameters. This yields $k=2.457$ $h=0.112$ $c = 8.69$. The CACC policy $\pi(u_i)$ is chosen as the relatively simple expression $u^{\text{ff}}_{i+1}=u_i$. For the MPC in~\cite{zheng2016distributed} we manually tuned the gains until adequate performance in the nominal case is reached. To highlight how the choice of parameters may affect the performance, we show the results for a time horizon of $N=20$ steps and for $N=40$ steps, where the simulation time discretization is $dt=0.05$s. Concerning~\cite{kafash2018constraining}, the method needs symmetric actuator bounds to work, so we set them equal to the maximum braking capacity, as this is a more conservative choice. In all simulations we consider that the platoon starts from steady state conditions, $v_i=v^D,\,\forall\,i\in\{1,...,n\}$ and $\Tilde{p}_i=0\,\forall\,i\in\{2,...,n\}$. Where $n$ is the number of vehicles in the platoon.

\begin{figure*}[t]
\centering
\includegraphics[width=\linewidth]{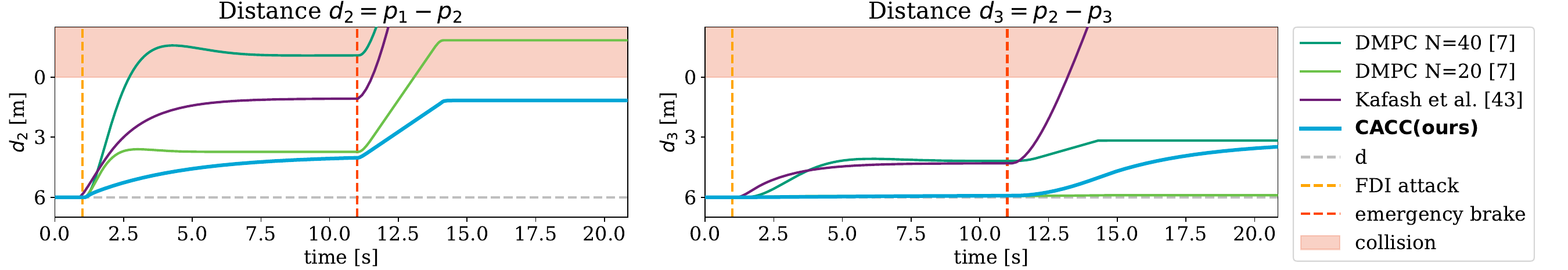}
    \caption{Comparison between different platooning controllers on a platoon of $3$ vehicles. From $t=1$s the leader is subject to an additive attack $u_0=u_0+u_{\min}$ that induces it to brake, while the first follower is subject to an additive attack equal to $u_1=u_1+u_{\max}$ that induces it to accelerate. The last vehicle is not affected by the attack. Additionally, at $t=11$s, the leader performs an emergency braking maneuver, i.e. $u_0=u_{\min}$. We show the distance between the leader and the first follower $d_2$ and between the first and second follower $d_3$. A collision occurs when the distance is less than $0$. }
  \label{fig:simulations illustrative example}
\end{figure*}

As an illustrative example we consider a platoon of $n=3$ vehicles where the leader is subject to an additive attack $u_0=u_0+u_{\min}$ that induces it to brake, while the first follower is subject to an additive attack equal to $u_1=u_1+u_{\max}$ that induces it to accelerate. Furthermore, after some time the leader performs an emergency brake, i.e. $u_0=u_{\min}$. This could happen if the leader encounters an obstacle on the road such as another human-driven traffic participant, or if it is affected by a more severe sensor-tampering attack such as Lidar spoofing~\cite{lidar_spoofing}. Figure~\ref{fig:simulations illustrative example} shows the distance between the first follower and the leader $d_2=p_1-p_2$ and between the second and first follower $d_3=p_2-p_3$ for the different methods.

To provide more statistical relevance to our comparison, we have conducted an extensive simulation study. We now consider a platoon of $n=11$ vehicles. We simultaneously subject all the vehicles in the platoon to an attack that replaces the true inbound acceleration signal from the predecessor with either a constant value $u_i(t) = c_i$, a sinusoidal signal defined as $u_i(t) = a_i\sin{(\phi_i+f_i2\pi t)}$ or a random value $u_i(t) = e_i \in [u_{\min},u_{\max}]$ taken from a uniform random distribution, which is then filtered using a first order filter with time constant $\tau_i$. For each method we run $1000$ simulations of $100$s for each attack type. We randomize the attack parameters for each vehicle in the platoon in every simulation run, ensuring that the attack signal remains within the actuation limits, i.e. $u\in [u_{\min},u_{\max}]$. Furthermore at $t=100$s the leader performs an emergency brake. The results are presented in table~\ref{tab:statistical_comparison_merged}.Note that the table shows the aggregated data for all vehicles across all simulation runs, since we did not notice any significant difference among vehicles.

By looking at the data we can see how the MPC in~\cite{zheng2016distributed} can be tuned to perform adequately for most attack types. However, since it does not have explicit safety guarantees, this process would require lengthily \textit{a posteriori} analysis. Interestingly, a shorter time horizon improves the safety margin both during the attack and during the emergency brake. This is because of the terminal set constraint in~\cite{zheng2016distributed}. Which forces the vehicle to reach the desired position at the end of the MPC horizon. A shorter horizon entails for a more reactive controller.

Concerning~\cite{kafash2018constraining}, we notice that during the attack phase of the simulation it performs well, outperforming its competitors in most cases. However, the emergency brake scenario falls outside the assumptions of the method, leading to collisions in our simulations. Indeed the method relies in bounding the communication-based contribution of the controller, while the sensor-based contribution is not bounded. In practice, however, they both contribute to the same actuation pool. An emergency brake can be considered as the result of an infinite negative additive attack that saturates the braking capabilities, violating the method's assumptions. Furthermore, computing the distance between vehicles when they have all reached a standstill, yields a solution of the form $d_{i}=d_{i+1} - \delta$, where $\delta>0$. This means that the inter-vehicle distance keeps decreasing moving from the tail to the head of the platoon. So for an arbitrarily long platoon a collision is unavoidable. 

Our method achieves $100\%$ safety under all tested attack conditions, including during emergency braking scenarios, while maintaining comparable performance to the baselines. These results validate the effectiveness of the safety filter introduced in Equation ~\ref{eq: uff}. While these findings are promising, we note that they are based on extensive but simulated scenarios. Real-world deployment may introduce additional uncertainties not captured in this study, which we aim to address in future work.

\begin{table}[h]
\centering
\begin{tabular}{c|c|c|c|c|c|c|}
\cline{2-7}
& \begin{tabular}[c]{@{}c@{}}mean\\dist.\\ {[}m{]}\end{tabular} 
& \begin{tabular}[c]{@{}c@{}}std.\\ dev.\\ {[}m{]}\end{tabular} 
& \begin{tabular}[c]{@{}c@{}}max\\dist.\\ {[}m{]}\end{tabular} 
& \begin{tabular}[c]{@{}c@{}}min\\dist.\\ {[}m{]}\end{tabular} 
& \begin{tabular}[c]{@{}c@{}}safe\\ runs\\ (attack)\end{tabular} 
& \begin{tabular}[c]{@{}c@{}}safe\\ runs\\ (brake)\end{tabular} \\ \hline

\multicolumn{7}{|c|}{Attack type: constant} \\ \hline
\multicolumn{1}{|c|}{DMPC N=40} & 6.07 & 4.25 & -2.79 & 44.50 & 91.35\% & 8.16\% \\ \hline
\multicolumn{1}{|c|}{DMPC N=20} & \textbf{6.00} & 1.31 & 3.76 & 8.21 & \textbf{100\%} & 53.98\% \\ \hline
\multicolumn{1}{|c|}{Kafash et al.} & 6.03 & \textbf{1.12} & 1.19 & 10.55 & \textbf{100\%} & 0\% \\ \hline
\multicolumn{1}{|c|}{CACC (ours)} & \textbf{6.01} & 1.15 & \textbf{4.00} & \textbf{7.98} & \textbf{100\%} & \textbf{100\%} \\ \hline

\multicolumn{7}{|c|}{Attack type: sinusoidal} \\ \hline
\multicolumn{1}{|c|}{DMPC N=40} & \textbf{6.00} & 0.78 & -2.79 & 33.03 & \textbf{100\%} & 0.01\% \\ \hline
\multicolumn{1}{|c|}{DMPC N=20} & 6.25 & 0.86 & 0.25 & 16.40 & \textbf{100\%} & 27.50\% \\ \hline
\multicolumn{1}{|c|}{Kafash et al.} & \textbf{6.00} & \textbf{0.26} & \textbf{4.93} & \textbf{7.27} & \textbf{100\%} & 0\% \\ \hline
\multicolumn{1}{|c|}{CACC (ours)} & \textbf{6.01} & 0.37 & 4.70 & \textbf{7.27} & \textbf{100\%} & \textbf{100\%} \\ \hline

\multicolumn{7}{|c|}{Attack type: random} \\ \hline
\multicolumn{1}{|c|}{DMPC N=40} & 6.04 & 1.17 & 0.02 & 15.70 & \textbf{100\%} & 0\% \\ \hline
\multicolumn{1}{|c|}{DMPC N=20} & 6.02 & 0.27 & 4.23 & 8.91 & \textbf{100\%} & 49.57\% \\ \hline
\multicolumn{1}{|c|}{Kafash et al.} & \textbf{6.00} & \textbf{0.13} & \textbf{5.17} & \textbf{6.81} & \textbf{100\%} & 0\% \\ \hline
\multicolumn{1}{|c|}{CACC (ours)} & 6.07 & 0.19 & \textbf{5.17} & 7.04 & \textbf{100\%} & \textbf{100\%} \\ \hline

\end{tabular}
\caption{Simulation study results under different attack types. For each method and for each attack type we performed $1000$ simulations of $100$ sec. For these simulations we disable the attack detection module to show the inherent robustness properties of the platooning controller. At the end of each simulation the leader performs an emergency brake. The mean represents the time-averaged value across all vehicles, and the standard deviation quantifies the temporal variation around this mean. Safe runs (attack) refers to the percentage of collisions over all vehicles when the attack is active, while safer runs (brake) refers to the collisions during the emergency brake. DMPC refers to~\cite{zheng2016distributed}, and Kafash et al. to~\cite{kafash2018constraining}.}
\label{tab:statistical_comparison_merged}
\end{table}

\section{Experimental results}
\label{sec:Experimental_results}
We tested our algorithm by using four scaled-down car-like robots shown in Figure~\ref{fig:jetracer intro picture}. We performed three experiments aimed at verifying different properties of the proposed platooning strategy: \emph{Experiment 1} shows the baseline platooning performance of the linear controller and the improvements that an additional feed-forward strategy can bring; \emph{Experiment 2} demonstrates the collision avoiding guarantees of the proposed strategy under a fake data injection attack; \emph{Experiment 3} shows the attack detection and mitigation strategies, as well as threat isolation by rearranging the platoon. Videos of the experiments are available at~\cite{video}. The details of the experimental set-up are described in the appendix. Concerning the communication protocol, we implemented the V2V communication layer over a standard wireless local area network (Wi-Fi) using ROS (Robot Operating System) messaging. While this setup does not replicate the exact physical and MAC-layer behavior of DSRC or C-V2X, it captures the essential characteristics of V2V communication relevant to our study—namely, message latency, packet loss, and the ability to inject or filter malicious data.

\subsubsection*{Experiment 1} 
The first experiment we conducted is aimed at highlighting the string stability property of the proposed sensor-based platooning algorithm and showing how a communication-based feed-forward term can improve performance.

\begin{figure*}[t]
	\centering
    \includegraphics[width=\textwidth]{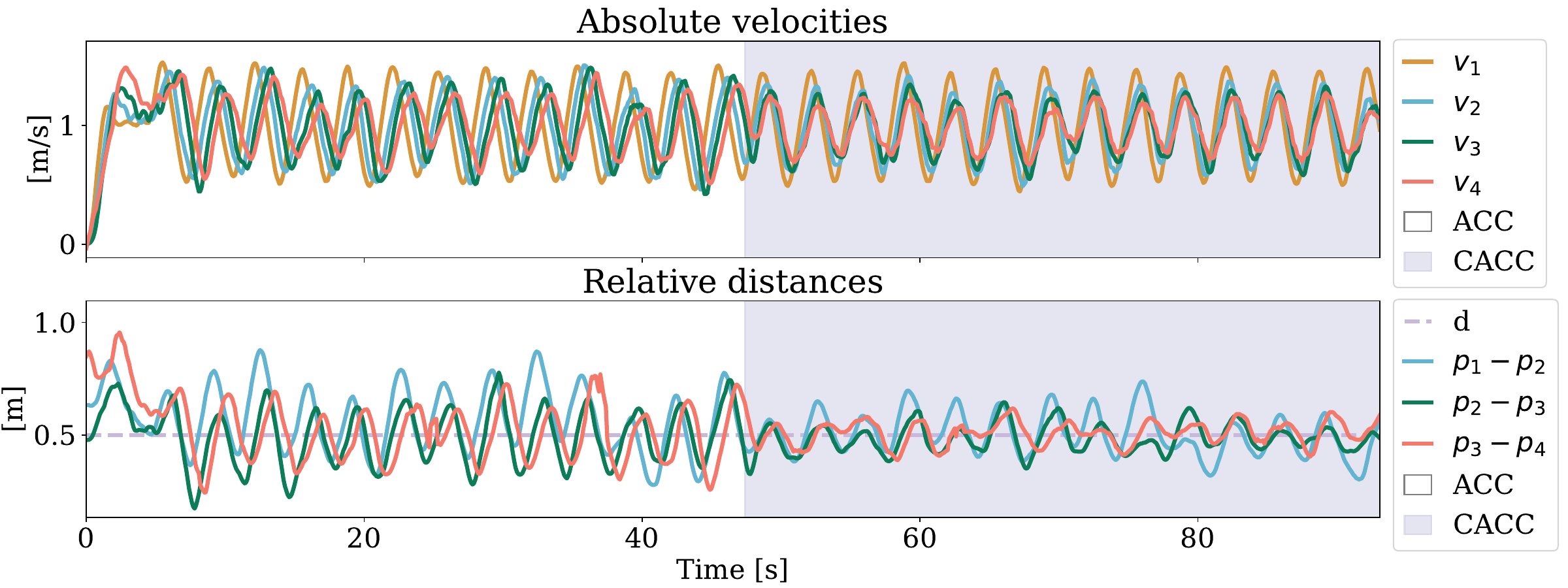}
	\caption{Experiment 1: the leader of the platoon follows a sinusoidal velocity reference, while other vehicles initially rely only on the linear sensor-based feedback controller. String stability properties are evident from the reduction in amplitude of the absolute velocity oscillations going towards the tail of the platoon. When the feed-forward action is activated (shaded area) the platooning performances increase, as can be seen in the reduction of relative position and velocity oscillations.}\label{fig:exp1}
\end{figure*}

We chose the linear controller gains as described in section~\ref{sec:platoon_controller}, with the following vehicle actuation limits: $u_{\max}=+1\text{m/s}^2, u_{\min}=-1\text{m/s}^2,v_{\max}=+1.4\text{m/s},v^D=1\text{m/s}$. These limits were set according to the hardware capabilities. We then selected a target inter-vehicle distance of $0.5$m and the lowest admissible value of $h$, see Figure~\ref{fig: gain tuning graph} for reference. The resulting gains are $k=3.45$, $h=0.21$ and $c=4.83$. The gain selection code is available on the GitHub repository~\cite{GitRepo}. We define the feed-forward policy simply as $u_{i+1}^{\text{ff}}(t) = u_{i}(t)$. To induce oscillations in the platoon and highlight the string stability properties we provide the leader with a sinusoidal velocity reference, initially without any feed-forward action. The experiment results are shown in Figure~\ref{fig:exp1}. Note that activating $u^{\text{ff}}$ should not result in perfect relative distance tracking, indeed $u^{\text{ff}}_{i+1}$ will compensate for the disturbance coming from $u_i$, but the external damping term $-kh(v_{i+1}-v^D)$ of the linear controller will still enforce string stability, damping out absolute velocity oscillations around $v_D$. If perfect relative distance tracking is desired, $u^{\text{ff}}_{i+1}$ should also compensate for the absolute damping term, i.e. $u^{\text{ff}}_{i+1}=u_{i+1}+kh(v_{i+1}-v^D)$, this can be seen from the simulations we performed in the GitHub repository ~\cite{GitRepo}. We also repeated the experiment and collected the distances data in table~\ref{tab:distances_sin_v}. The string stability property of both algorithms can be seen from the reducing standard deviation values along the platoon, e.g. Vehicle 3 has smaller std dev than Vehicle 2. The performance increase when activating CACC mode is highlighted bold. The mean gets closer to $d=0.5$m, the std dev reduces and the min and max value get closer to $d=0.5$m. Notice that "Vehicle $i+1$" refers to the distance $p_i-p_{i+1}$.
\begin{table}[h]
\begin{tabular}{c|cc|cc|cc|}
\cline{2-7}
                                      & \multicolumn{2}{c|}{Vehicle 2}                                    & \multicolumn{2}{c|}{Vehicle 3}                                    & \multicolumn{2}{c|}{Vehicle 4}                                    \\ \cline{2-7}
\multicolumn{1}{l|}{}                 & \multicolumn{1}{l|}{ACC} & \multicolumn{1}{l|}{{CACC}} & \multicolumn{1}{l|}{ACC} & \multicolumn{1}{l|}{{CACC}} & \multicolumn{1}{l|}{ACC} & \multicolumn{1}{l|}{{CACC}} \\ \hline
\multicolumn{1}{|c|}{mean {[}m{]}} & \multicolumn{1}{c|}{0.45} & \textbf{0.46} & \multicolumn{1}{c|}{0.52} & \textbf{0.51} & \multicolumn{1}{c|}{0.53} & \textbf{0.51} \\ \hline
\multicolumn{1}{|c|}{std dev {[}m{]}} & \multicolumn{1}{c|}{0.09} & \textbf{0.06} & \multicolumn{1}{c|}{0.06} & \textbf{0.04} & \multicolumn{1}{c|}{0.06} & \textbf{0.04} \\ \hline
\multicolumn{1}{|c|}{max {[}m{]}} & \multicolumn{1}{c|}{\textbf{0.64}} & 0.67 & \multicolumn{1}{c|}{0.69} & \textbf{0.63} & \multicolumn{1}{c|}{0.74} & \textbf{0.67} \\ \hline
\multicolumn{1}{|c|}{min {[}m{]}} & \multicolumn{1}{c|}{0.25} & \textbf{0.30} & \multicolumn{1}{c|}{0.37} & \textbf{0.40} & \multicolumn{1}{c|}{0.38} & \textbf{0.41} \\ \hline
\cline{2-7}
\end{tabular}
\caption{Distances for ACC and CACC platooning with leader executing a sinusoidal velocity profile. Notice that "Vehicle $i+1$" refers to the distance $p_i-p_{i+1}$. The data is collected over a $60$s period for each control algorithm. The mean represents the time-averaged value, and the standard deviation quantifies the temporal variation around this mean.}
\label{tab:distances_sin_v}
\end{table}

\subsubsection*{Experiment 2}
\begin{figure*}[t]
	\centering
	\includegraphics[width=\textwidth]{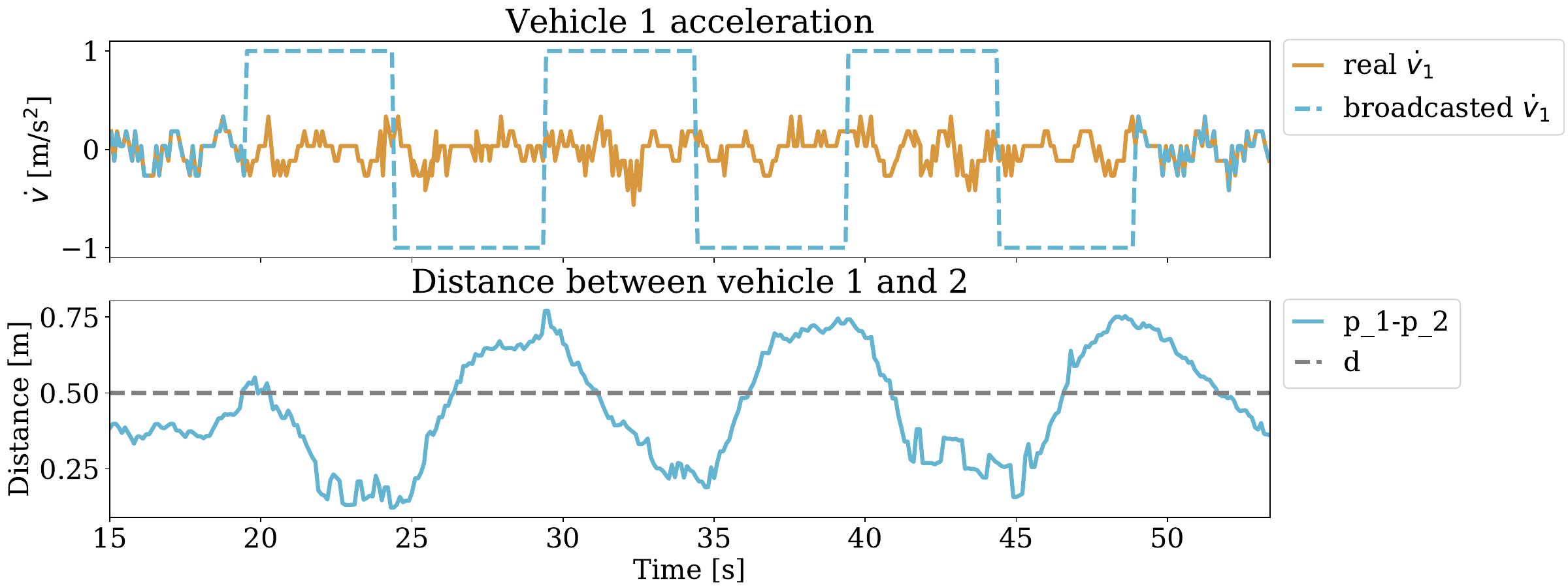}
    \caption{Experiment 2: We simulate a fake data injection attack between the leader and vehicle $2$. The collision avoidance properties of the proposed algorithm prevent vehicle $2$ from crashing into the leader. Without any attack detection module however, the attacker is still able to manipulate the distance between the vehicles.}\label{fig:exp2}
\end{figure*}

The second experiment shows the collision avoiding properties of the proposed algorithm in the presence of an undetected fake data injection attack. While $u^{\text{ff}}$ is activated we simulate a fake data injection attack between the leader and the second vehicle in the platoon and disable the attack detection module. Instead of the real $u_1$ value the leader communicates alternating values $u_1(t)=u_{\max}$ or $u_1(t)=u_{\min}$ every $5$s. The experiment results are shown in Figure~\ref{fig:exp2}. Collision avoidance properties are visible from the fact that even when the leader communicates $u_1(t)=u_{\max}$ and vehicle $2$ is tricked into accelerating and gets closer, no collision occurs thanks to constraint~\eqref{eq: uff_no_sat_lin} being activated, that is, the relative distance remains $p_i-p_{i+1}\geq 0$. However, we notice that without an attack detection module the attacker is still able to manipulate the distance, degrading platooning performance.

\subsubsection*{Experiment 3} In the last experiment we introduce the attack detection module and show the attack mitigation strategies of the proposed algorithm, namely the disabling of the compromised $u_i$ channel and the platoon rearrangement. 

To select appropriate values of $K$ and $\Bar{r}$ for the attack detection module we used data from \emph{Experiment 2} and simulated the residual dynamics according to different $K$ values and chose corresponding appropriate $\Bar{r}$ threshold values. For the current experiment we chose $K=0.05$ and $\Bar{r}=0.75$.
Note that properly tuning the low-level longitudinal controller in equation~\eqref{eq:low-level controller} in the appendix will strongly affect the ability of the vehicles to accurately track the $u_i(t)$ and $u_{i+1}(t)$ profiles outputted by the platooning controller. Poor tracking performance will thus hinder the attack detection module, as the effect of a malicious attack will be "hidden" by the baseline process noise. To further validate our tuning process we also recorded data during nominal platooning conditions and simulated different kinds of attacks, namely by replacing the recorded $u_1$ with a sinusoidal signal, and by adding a Gaussian noise, both with different sets of parameters such as frequency and intensity. The healthy system data and the attack simulation code is available on the GitHub repository~\cite{GitRepo}.

\begin{figure*}[t]
	\centering
	\includegraphics[width=\textwidth]{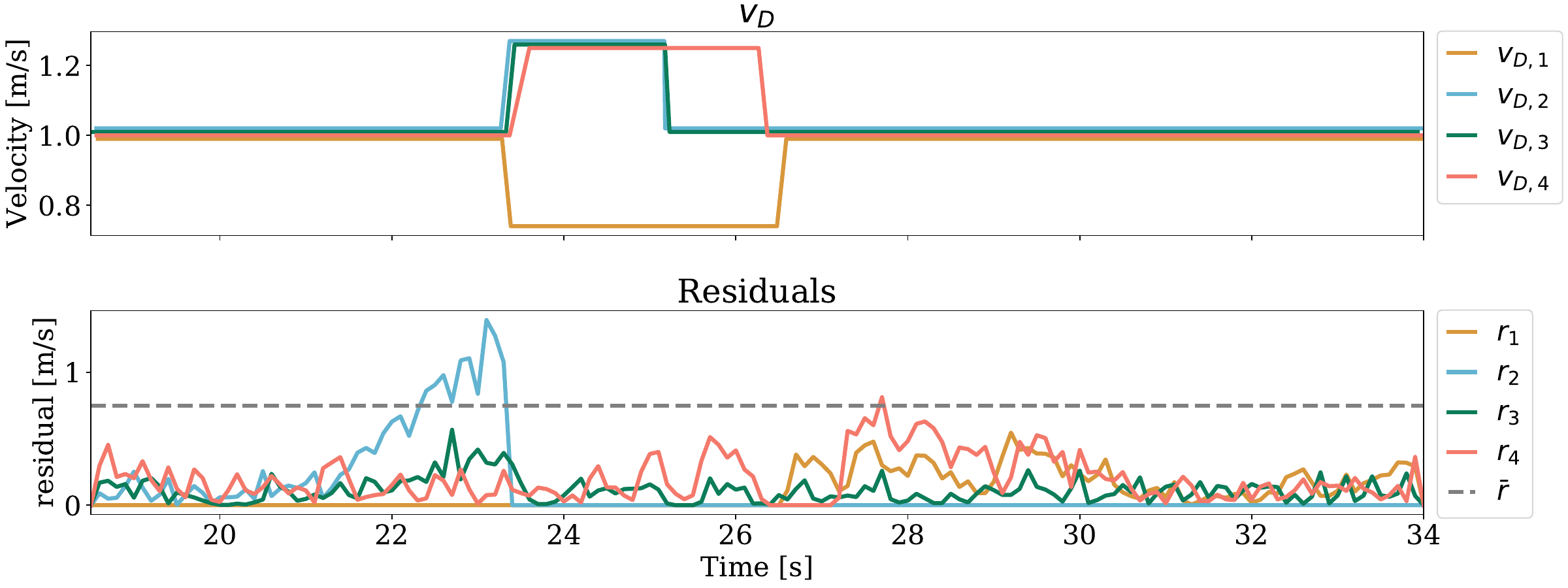}
	\caption{Experiment 3: We show the attack mitigation properties of the proposed platooning strategy by activating the attack detection module and subjecting the system to the same fake data injection attack as in Experiment 2. The attack is detected within a few seconds , as can be seen from the sudden drop in $r_2$ after it exceeds the threshold, signaling that the communication-based $u^{\text{ff}}_2$ has been disabled. This stops the attack from taking control of vehicle $2$ as in Experiment 2. The platoon coordinator module then changes the topology and assigns the compromised vehicle to the end of the platoon. The \textit{platoon rearranging logic}, detailed in appendix~\ref{sec:appendix}, then executes the overtaking maneuver. Vehicles $2,3,4$ switch to the overtaking lane, while vehicle $1$ slows down to facilitate the overtaking. Once all vehicles have merged back into the slow lane and the new topology has been achieved, the compromised vehicle has no follower to communicate with. The attack has thus been isolated and the platoon completely recovers nominal functionality. Notice that the platoon leader (first vehicle 1 and then vehicle 2 after the topology change) has no inbound communication to assess, to indicate this we set $r=0$. For visual clarity we also offset the $v_D$ by a small amount.}\label{fig:exp3}
\end{figure*}

We now demonstrate the attack mitigation properties of the proposed algorithm by enabling the attack detector module and subjecting the platoon to the same fake data injection attack as in \emph{Experiment 2}. As shown in Figure~\ref{fig:exp3}, the attack is detected within few seconds, note that for this experiment we additionally require that $r>\Bar{r}$ for $0.5$s consecutively to consider the system under attack (and set $\sigma = 0$) as measure to reduce false positives. Since while the attack is detected $u^{\text{ff}}$ is disabled, the attack no longer has the effects visible in \emph{Experiment 2}. Note that we disable the feed-forward action only on vehicle $2$, the rest of the platoon maintains nominal functionality. The Platoon Coordinator onboard each vehicle then changes the platoon topology and the compromised vehicle is assigned to the end of the platoon. To perform this maneuver vehicles $2,3,4$ switch to an overtaking lane and then switch back to the original platooning lane as shown in the video of the experiments~\cite{video}. To execute the platoon rearranging in practice we have designed a fully distributed logic that runs onboard each vehicle. The outputs of this logic are mainly the lane the vehicle should be in, i.e. the slow lane or the overtaking lane, and the velocity reference $v_D$. The latter is needed to speed up the vehicle once it is in the overtaking lane, or to slow it down when waiting to be overtaken by other vehicles. Details of the platoon rearranging logic are discussed in appendix~\ref{sec:appendix}.

As shown in Figure~\ref{fig:exp3}, after the new topology has been reached, since the compromised vehicle has no follower to communicate with, the attack has been isolated and the platoon recovers complete nominal functionality, i.e. all the residuals are below the threshold. Notice that vehicle $2$ is the new platoon leader and thus has no inbound communication, to indicate this we set $r_2=0$.

\section{Conclusions}
\label{sec:Conclusions}

We present our distributed attack-resilient platooning strategy. We combine a safe and string stable platoon controller with a platoon coordinator, which is able to reorganize the platoon. Our proposed method preserves string stability and safety guarantees in case of a false acceleration data attack, as supported by the simulation and experimental results. Future works may include considering other communication topologies between vehicles and other attacker models, working on more sophisticated trustworthiness estimation algorithms, and testing our algorithm with an increased number of robots.
\appendix
\section{Appendix}\label{sec:appendix}
\emph{Experimental set-up.} For the experiments we used DART, the robotic platform presented in our previous work~\cite{DART}. The robots are equipped with a wheel encoder to measure their own speed and a Lidar to perceive the surrounding environment. The team of robots follows a predefined path while the platooning controller regulates the longitudinal distances. Each robot runs the perception pipeline and control algorithms on-board, i.e. the platooning strategy runs in a fully distributed fashion. Communication among robots is handled by a dedicated ROS network. The code used during the experiments can be found in the publicly available GitHub repository~\cite{GitRepo}. 

\begin{figure}[t]
\centering
\includegraphics[width=\columnwidth]{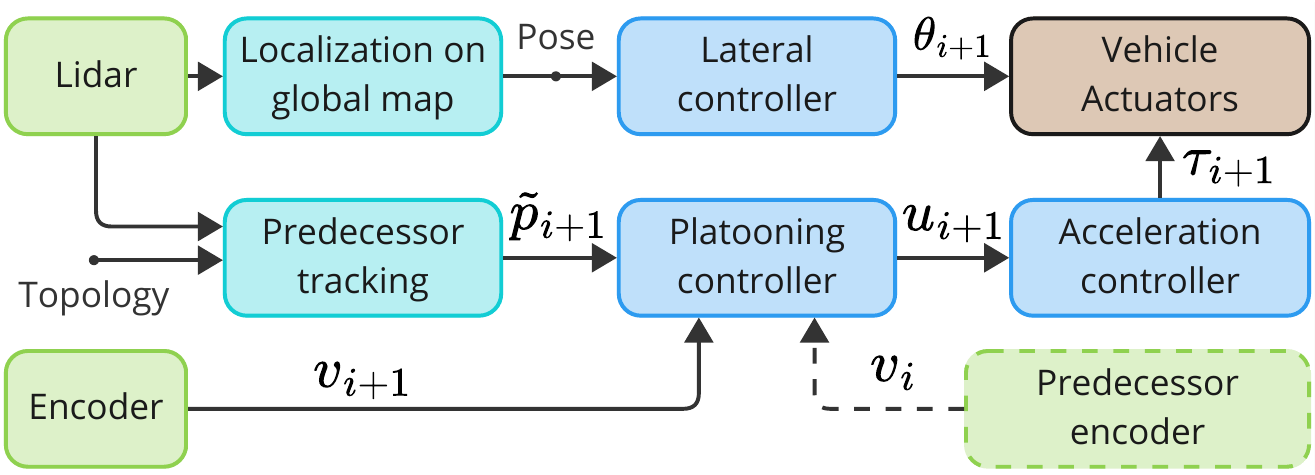}
    \caption{Overview of the processes running on-board the $(i+1)$-th vehicle.}
  \label{fig:inner_workings}
\end{figure}

The proposed platooning algorithm is designed considering the vehicles' longitudinal dynamics as a linear system that can be controlled with acceleration inputs. To run the proposed platooning algorithm on real robots we must first design three additional modules, namely: a lateral controller to follow the predefined path, a perception pipeline to measure the relative distance and velocity from the preceding vehicle, and a low level controller that enforces the longitudinal acceleration commands coming from the platooning controller. Figure~\ref{fig:inner_workings} shows an overview of the processes running on board each robot. We will now address each component.

\emph{Lateral controller.} To replicate a highway scenario the predefined path features straight sections and gentle curves. Under these circumstances the lateral and longitudinal vehicle dynamics can be decoupled, allowing us to design the lateral controller separately from the platooning controller. To perform path tracking we rely on the standard ROS libraries to estimate each robot's pose and position on a previously built map. The steering angle $\theta_{i+1}$ is then evaluated using a pursuit controller~\cite{pursuit_controller}. Note that each robot follows the path independently, circumventing any possible lateral string stability issues.

\emph{Perception pipeline.} The longitudinal controller requires three inputs, namely the $(i+1)$-th vehicle velocity $v_{i+1}$, and the relative velocity and position of the preceding vehicle, $\Tilde{v}_{i+1}$ and $\Tilde{p}_{i+1}$ respectively. $v_{i+1}$ is readily available since each vehicle is equipped with a wheel speed encoder. To evaluate $\Tilde{v}_{i+1}$ a full scale vehicle would typically rely on sensors such as frequency-modulated RADAR and Depth cameras, yet due to the limited sensing capabilities of the platform we use communication, i.e. we evaluate $\Tilde{v}_{i+1}$ as ${v}_{i+1}-{v}_{i}$ where ${v}_{i}$ is communicated by the $i$-th vehicle. The relative distance $\Tilde{p}_{i+1}$ is measured with the on-board Lidar. Each vehicle processes the incoming $2$D Lidar scan and extracts clusters of points that represent different objects. The robots choose the vehicle to follow (VTF) as detailed in the \textit{Platoon rearranging logic} at the end of this section. The distance to the vehicle to follow is then measured as the distance to the centre of the cluster of points closest to the VTF's advertised position on the global map. This can be seen in the experiments video~\cite{video}. 

\emph{Low-level acceleration tracking controller.} The platooning controller receives $v_{i+1}$, $\Tilde{v}_{i+1}$, $\Tilde{p}_{i+1}$ and produces a control action $u_{i+1}$ in $\frac{m}{s^2}$, yet the vehicle's actuators only accept throttle values $\tau\in[0,1]$. We thus need a low-level controller that tracks $u_{i+1}$. We designed the low-level longitudinal controller as:
\begin{align}\label{eq:low-level controller}
    \tau_{i+1}=\tau_{i+1}^{\text{ff}}(u_{i+1},v_{i+1}) + \tau_{i+1}^{\text{fb}}.
\end{align}
Where $\tau_{i+1}^{\text{ff}}(u_{i+1},v_{i+1})$ is a feed-forward term obtained by solving the inverse system dynamics, i.e. by inverting the dynamic model $u=f(\tau,v)$. To obtain a reliable dynamic model of the vehicles $f$ we relied on the system identification procedure presented in~\cite{DART}. The feedback term $\tau_{i+1}^{\text{fb}}$ is an additional corrective integral action that is needed to compensate for differences among robots and slowly time-varying phenomena such as the electric batteries depleting and subsequently changing the supplied voltage to the driving motors. Note the design and tuning of the low level controller strongly affects the overall performance of the platooning controller, since it will affect the resulting acceleration profile. It may even affect the string stability property if the requested $u_{i+1}$ is tracked with some time delay. This is why we advise against using a Proportional-Derivative structure in the low level controller and suggest using small gains for the integral action $\tau_{i+1}^{\text{fb}}$.\\
\emph{Platoon rearranging logic.}
In \textit{Experiment 3} once an attack is detected and one of the communication channels is closed, the platoon coordinator can isolate the compromised vehicle by moving it to the end of the platoon. To perform this rearranging maneuver in practice, we have designed a fully distributed logic that runs onboard each vehicle. The outputs of this logic are which lane the vehicle should follow, its velocity reference $v_D$ and what vehicle it should follow. The latter is mainly needed to avoid abrupt changes in velocity since when a new desired topology is outputted by the platoon coordinator, the $i$-th vehicle's assigned predecessor may not be in sight yet. The inputs to the logic are: \emph{1)} the desired topology coming from the platoon coordinator (that also runs on each vehicle in a distributed fashion) that indicates the assigned predecessor (AP), \emph{2)} the observed predecessor (OP), that is the closest vehicle in front of the ego vehicle regardless of the lane it is in, \emph{3)} the lane (L) the ego vehicle is currently in, L=Slow Lane (L=SL) or L=Fast Lane (L=FL), \emph{4)} the lane its assigned predecessor is currently in (AP L), \emph{5)} if its assigned predecessor is signaling with the right indicator, i.e. if it intends to merge into the slow lane (MSL). When a vehicle is in the slow lane (L=SL) and it's observed predecessor is its assigned predecessor, i.e. L=SL and AP=OP, the ego vehicle has reached the correct platooning position (CPP). Notice that AP=$0$ indicates the ego vehicle should be the leader of the platoon. On a real full-scale vehicle all the required inputs can be gathered using cameras, since all the data concerns the immediate surroundings of the ego vehicle. For the experiments we rely on the advertised position of each vehicle on the global map and on communication, for example to detect if MSL is active or not. 

\begin{figure}[htbp]
  \centering
  \begin{minipage}[b]{0.24\linewidth}
    \includegraphics[width=\linewidth]{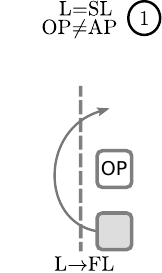}
  \end{minipage}
  \begin{minipage}[b]{0.24\linewidth}
    \includegraphics[width=\linewidth]{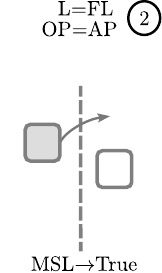}
  \end{minipage}
  \begin{minipage}[b]{0.24\linewidth}
    \includegraphics[width=\linewidth]{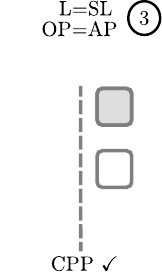}
  \end{minipage}

  \caption{Leader lane management logic. The ego vehicle is represented as a shaded rectangle.}
  \label{fig: leader lane manager}
\end{figure}

\begin{figure}[htbp]
  \centering
  \begin{minipage}[b]{0.24\linewidth}
    \includegraphics[width=\linewidth]{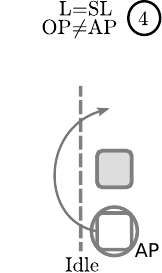}
  \end{minipage}
  \begin{minipage}[b]{0.24\linewidth}
    \includegraphics[width=\linewidth]{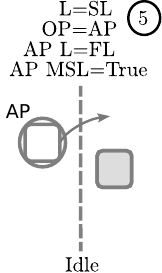}
  \end{minipage}
  \begin{minipage}[b]{0.24\linewidth}
    \includegraphics[width=\linewidth]{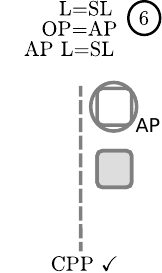}
  \end{minipage}

  \vspace{1 em} 

  \begin{minipage}[b]{0.24\linewidth}
    \includegraphics[width=\linewidth]{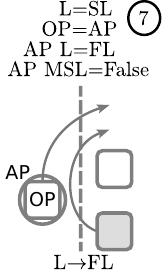}
  \end{minipage}
  \begin{minipage}[b]{0.24\linewidth}
    \includegraphics[width=\linewidth]{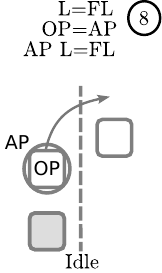}
  \end{minipage}
  \begin{minipage}[b]{0.24\linewidth}
    \includegraphics[width=\linewidth]{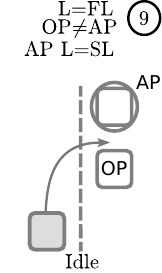}
  \end{minipage}
  \begin{minipage}[b]{0.24\linewidth}
    \includegraphics[width=\linewidth]{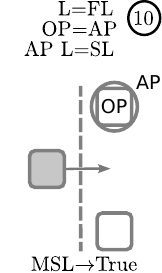}
  \end{minipage}

  \caption{Follower lane management logic. The ego vehicle is represented as a shaded rectangle.}
  \label{fig: follower lane logic}
\end{figure}

We define three separate logic controllers, each regulating the lane choice, $v_D$ and vehicle to follow respectively.
\textit{Lane manager.} To determine in which lane the vehicle should be in, we have identified $10$ different scenarios the vehicles may encounter and programmed the required actions. We first differentiate if the ego vehicle is the new leader or not, if it is ($\text{AP}=0$) it will encounter scenarios $1$-$3$, shown in figure~\ref{fig: leader lane manager}. If not ($\text{AP}\neq0$) it will encounter scenarios $4$-$10$ shown in figure~\ref{fig: follower lane logic}. When changing lanes, merging to the fast lane is always allowed because in these experiments we don't consider other traffic participants that may be overtaking. When merging to the slow lane instead we must check that there is enough space between the ego vehicle and the following vehicle. In this case MSL=True but the vehicle is still in the fast lane L=FL. The information about MSL is also necessary to disambiguate scenario 5 from scenario 7 that would otherwise be identical. \textit{Velocity manager.} We apply a simple logic to regulate the speed of the ego vehicle in order to facilitate overtaking, shown in algorithm~\ref{algorithm: vd manager}.
\textit{Vehicle to follow manager.} Vehicle to follow (VTF) refers to which vehicle the ego vehicle tries to keep the desired distance $d$ from. To avoid abrupt changes the logic shown in algorithm~\ref{algorithm: vehicle to follow manager} enables to switch only if the assigned predecessor is in front of the ego vehicle.

\begin{algorithm}
\caption{Decision logic for $v_D$}
\begin{algorithmic}
\If{$L = SL$}
    \If{$AP = OP$}
        \State $v_D \gets \text{default}$
    \Else
        \State $v_D \gets \text{slow}$
    \EndIf
\Else
    \If{$AP L = SL$}
        \State $v_D \gets \text{default}$
    \Else
        \State $v_D \gets \text{fast}$
    \EndIf
\EndIf
\end{algorithmic}
\label{algorithm: vd manager}
\end{algorithm}

\begin{algorithm}
\caption{Decision logic for vehicle to follow}
\begin{algorithmic}
\If{$\text{AP} = \text{OP}$}
    \State $\text{VTF} \gets \text{AP}$
\Else
    \If{L=SL}
        \State $\text{VTF} \gets 0$
    \Else
        \State $\text{VTF} \gets \text{closest preceding vehicle in FL}$ 
    \EndIf
\EndIf
\end{algorithmic}
\label{algorithm: vehicle to follow manager}
\end{algorithm}

\bibliographystyle{IEEEtran}
\bibliography{reference}

\begin{IEEEbiography}[{\includegraphics[width=1in,height=1.33in,clip,
    keepaspectratio]{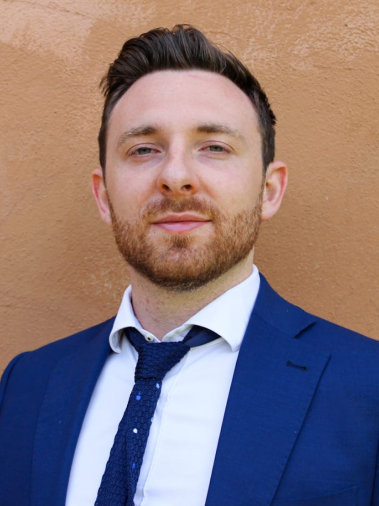}}]{Lorenzo Lyons} received the M.Sc. degree in
mechanical engineering from the Polytechnic University of Milan, Milan, Italy, in 2021. He is currently pursuing the Ph.D. degree with the Cognitive
Robotics (CoR) Department, Delft University of
Technology, Delft, The Netherlands.
His research interests include numerical optimization, model predictive control, multi-robot motion
planning applied to the automotive, and robotics.
\end{IEEEbiography}
\vspace{-\baselineskip}
\begin{IEEEbiography}[{\includegraphics[width=1in,height=1.33in,clip,
    keepaspectratio]{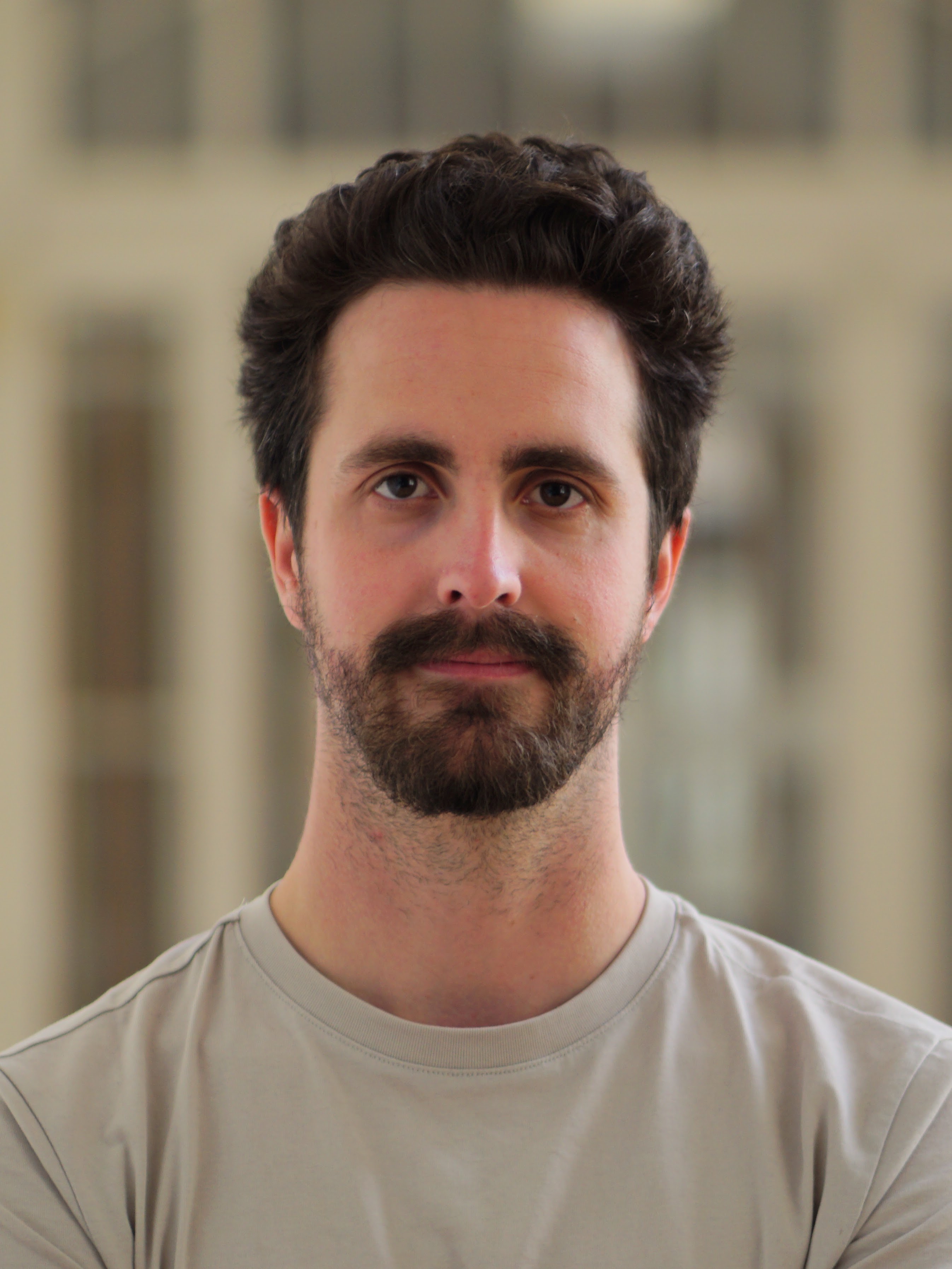}}]{Manuel Boldrer} received the Ph.D in Materials, Mechatronics and Systems Engineering from the University of Trento, Trento, Italy in 2022. He was a Visiting Scholar at the University of California, Riverside, Riverside, US, in 2021. He was a Postdoctoral Researcher with the Cognitive Robotics Department (CoR), Delft University of Technology, Delft, The Netherlands, in the Reliable Robot Control Lab from 2022 to 2023. As of today, he is a Researcher at the Multi Robot System group (MRS), Czech Technical University, Prague, Czechia. His research interests include mobile robotics, distributed
  control and multi-agent systems.
\end{IEEEbiography}
\vspace{-\baselineskip}
\begin{IEEEbiography}[{\includegraphics[width=1in,height=1.33in,clip,
    keepaspectratio]{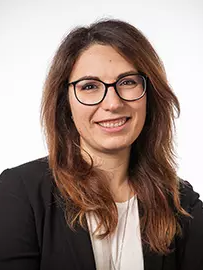}}]{Laura Ferranti} received the Ph.D. degree
from the Delft University of Technology, Delft,
The Netherlands, in 2017.
She is currently an Assistant Professor with
the Cognitive Robotics (CoR) Department,
Delft University of Technology. Her research
interests include optimization and optimal
control, model predictive control, reinforcement
learning, embedded optimization-based control with
application in flight control, maritime transportation,
robotics, and automotive.
Dr. Ferranti was a recipient of the NWO Veni Grant from the Netherlands
Organization for Scientific Research in 2020 and the Best Paper Award
on Multi-Robot Systems at International Conference on Robotics and
Automation (ICRA) 2019.
\end{IEEEbiography}
\vspace{-\baselineskip}

\end{document}